\documentclass[11pt]{article}
\usepackage[affil-it]{authblk}

\usepackage{color}
\usepackage{amsmath,amsfonts,amssymb,mathrsfs,amsthm}
\usepackage{graphicx}
\usepackage{listings}
\usepackage{enumerate,enumitem}
\usepackage{mathtools}
\usepackage[a4paper, total={6in, 8in}]{geometry}
\usepackage{cite}
\usepackage[colorlinks,linkcolor=blue,citecolor=blue]{hyperref}
\usepackage{empheq}
\usepackage{pdfpages}
\usepackage[titletoc]{appendix}
\usepackage[utf8]{inputenc}
\usepackage{xspace}
\numberwithin{equation}{section}

\newcommand{\Eqref}[1]{Eq.~\eqref{#1}}

\newcommand{\subfig}[2]{ Fig.~\hyperref[#1]{\ref{#1}#2}}

\newcommand{\pa}{\partial}

\newcommand{\be}{\begin{equation}}
	\newcommand{\ee}{\end{equation}}

\newcounter{mnotecount}[section]
\let\oldmarginpar\marginpar
\renewcommand\marginpar[1]{\-\oldmarginpar[\raggedleft\footnotesize #1]%
	{\raggedright\footnotesize #1}}

\title{Turing and wave instabilities in hyperbolic reaction-diffusion systems: The role of second-order time derivatives and cross-diffusion terms on pattern formation}
\author[1]{Joshua S. Ritchie\footnote{Email: jritchie@maths.otago.ac.nz.} }
\author[2]{Andrew L. Krause}
\author[1]{Robert A. Van Gorder\footnote{Email: rvangorder@maths.otago.ac.nz.} }
\affil[1]{Department of Mathematics and Statistics, University of Otago,  New Zealand.}
\affil[2]{Department of Mathematical Sciences, Durham University, Upper Mountjoy Campus, Stockton Rd, Durham DH1 3LE, United Kingdom.}
\date{\today}

\begin{document}
	\maketitle
	
	
	\begin{abstract}
		Hyperbolic reaction-diffusion equations have recently attracted attention both for their application to a variety of biological and chemical phenomena, and for their distinct features in terms of propagation speed and novel instabilities not present in classical two-species reaction-diffusion systems. We explore the onset of diffusive instabilities and resulting pattern formation for such systems. Starting with a rather general formulation of the problem, we obtain necessary and sufficient conditions for the Turing and wave instabilities in such systems, thereby classifying parameter spaces for which these diffusive instabilities occur. We find that the additional temporal terms do not strongly modify the Turing patterns which form or parameters which admit them, but only their regions of existence. This is in contrast to the case of additional space derivatives, where past work has shown that resulting patterned structures are sensitive to second-order cross-diffusion and first-order advection. We also show that additional temporal terms are necessary for the emergence of spatiotemporal patterns under the wave instability. We find that such wave instabilities exist for parameters which are mutually exclusive to those parameters leading to stationary Turing patterns. This implies that wave instabilities may occur in cases where the activator diffuses faster than the inhibitor, leading to routes to spatial symmetry breaking in reaction-diffusion systems which are distinct from the well studied Turing case.
	\end{abstract}

	\section{Introduction}
	
	Turing first described the emergence of spatial instability and resulting pattern formation in two coupled reaction-diffusion systems in \cite{Turing1952}, and since that time, the eponymous Turing instability has become a popular mechanism for investigating pattern formation in biological, chemical, and physical systems \cite{ouyang1991transition, lengyel1992chemical, seul1995domain, maini1997spatial, maini2006turing, baker2008partial, marcon2012turing}. Among the many generalizations to these so-called ``Turing systems", hyperbolic reaction-diffusion systems (reaction-diffusion systems where one or more of the equations also involves the second time derivative of the unknown field) have recently attracted interest for applications to ecology in the form of vegetation models for semiarid environments \cite{consolo2017pattern, consolo2019supercritical}, to models of movement and aggregation \cite{eftimie2012hyperbolic}, to epidemic models \cite{barbera2013spread}, to interacting cell systems relevant to ripple formation of myxobacteria \cite{lutscher2002emerging} and to the Brusselator autocatalytic reaction \cite{cho1993hyperbolic, al1996hyperbolic}. Among other applications, we mention two physical justifications for employing hyperbolic reaction-diffusion equations. The first is in considering models of time-delayed reaction-diffusion systems. One can then treat the time delay, $\tau$, as a small parameter and expand the fields in $\tau$, retaining up to the $\mathcal{O}(\tau)$ terms. Time delays are relevant in models of gene expression and cellular biology \cite{gaffney2006gene, lee2010influence} and also predator-prey dynamics \cite{xu2015pattern}. Hence, hyperbolic reaction-diffusion systems are of relevance to understanding the impact of such delays, without having to resolve delayed partial differential equations which are numerically and analytically cumbersome. A second distinct physical justification is capturing finite propagation speed associated with velocity-jump processes, rather than the infinite speed of propagation associated with classical Brownian motion. Many different formalisms exist for the latter case, with hyperbolic reaction-diffusion equations being the simplest, although not necessarily the most physical \cite{mendez2010reaction}. 
	
	Zemskov and Horsthemke \cite{zemskov2016diffusive} studied hyperbolic reaction-diffusion systems of the form 
	\begin{equation}\begin{aligned}\label{oldpaper}
			\tau\frac{\pa^2 u}{\pa t^2} + \frac{\pa u}{\pa t} & =  d_{11}\frac{\pa^2 u}{\pa x^2} +f\left(u,v\right),\\
			\tau\frac{\pa^2 v}{\pa t^2} + \frac{\pa v}{\pa t} & =  d_{22} \frac{\pa^2 v}{\pa x^2} +g\left(u,v\right).
	\end{aligned}\end{equation}
	They show that such equations admit the Turing instability or the wave instability under specific algebraic conditions on the parameters and spatial spectrum. In particular, given a spatial perturbation of the constant steady state $u^*,v^*$, they provide conditions on the model parameter and wavenumber of the perturbation which leads to one of these two instabilities. They demonstrate the occurrence of both instabilities in several specific systems. The framework and analysis of \cite{zemskov2016diffusive} was later extended to include superdiffusion (fractional diffusion $\nabla^{\alpha}$ with $1\leq \alpha \leq 2$) by \cite{mvogo2018diffusive}. Numerical simulations of related models with fractional diffusion were recently given in \cite{macias2019numerical, macias2019algorithm}. Hyperbolic systems with cross-diffusion terms were recently studied in \cite{curro2021pattern}, where it was pointed out the Turing and wave instabilities take place in distinct regions of parameter space for the model studied. See also further work of those authors in \cite{consolo2022oscillatory}. It is also possible to study hyperbolic reaction-diffusion systems on discrete or network domains, and inertial parameters were shown to induce oscillations in the network dynamics in \cite{carletti2021turing} as well as continuum dynamics in \cite{ghorai2022diffusive}.
	
	Motivated by the aforementioned applications to pattern formation, we classify the Turing and wave instabilities in general hyperbolic reaction-diffusion systems. We first outline the generic hyperbolic reaction-diffusion system in Sec. \ref{secmodel}, where we also provide a greater physical motivation for these systems. We then carry out the theoretical work in Sec. \ref{sectheory}, deriving conditions for the occurrence of Turing and wave instabilities which lead, respectively, to spatial and spatiotemporal patterns within hyperbolic reaction-diffusion systems of the form \eqref{Eq:GeneralProblem}. To illustrate the theory, we carry out fully nonlinear numerical simulations of hyperbolic reaction-diffusion systems in Sec. \ref{secexamples}. We find that Turing patterns obtained are qualitatively the same under hyperbolic and classical systems, whereas spatiotemporal patterns emerging from the wave instability are quite sensitive to the additional temporal terms. We briefly discuss our results and highlight key findings in Sec. \ref{secdisc}. We list specific conclusions resulting from our work in Sec. \ref{secconclusions}.

	\section{Hyperbolic reaction-diffusion systems}\label{secmodel}
	Standard reaction-diffusion systems are at best approximations to the mean-field behavior of particles, and contain many features which are not realistic in many physical applications. A key example of such features is the infinite propagation speed of disturbances in these parabolic systems \cite{mendez2010reaction}. One phenomenological way of addressing the issue with propagation speed is to consider a hyperbolic reaction-diffusion system \eqref{oldpaper}, which can be shown to admit both finite speeds of propagation, and behavior comparable to the standard case in the singular limit of $\tau \to 0$, at least for dissipative dynamics tending toward an equilibrium state. Another approach, which we outline below, reconsiders the derivation of a reaction-transport system via conservation of mass and a modified description of the flux \cite{joseph1989heat}. The conservation of mass equation reads 
	\begin{align}
		\frac{du}{dt} =  -\nabla \cdot \mathbf{j} + f(u),
		\label{Eq:ConservationMass} 
	\end{align}
	where $\mathbf{j}$ is the flux of the concentration $u$ and $f$ is the function denoting the reaction kinetics. The Cattaneo equation gives a deviation from Fickian diffusion which accounts for a finite relaxation time of particle motion, and is given by
	\begin{align}
		\tau \frac{\partial \mathbf{j}}{\partial t} + \mathbf{j} = -d \nabla u,
		\label{Eq:CattaneoEquation}
	\end{align}
	where $\tau,d>0$ are both constants. In the special case $\tau=0$, \eqref{Eq:CattaneoEquation} reduces to Fick's law. Eq. \eqref{Eq:CattaneoEquation} therefore generalizes Fick's law so that the flux $\mathbf{j}$ does \emph{not} adjust instantaneously to the gradient of $u$. Instead there is a relaxation time $\tau$. Taking the derivative of  \eqref{Eq:ConservationMass} with respect to $t$, and the divergence of  \eqref{Eq:CattaneoEquation}, we can combine these equations (by eliminating mixed derivatives) and arrive at the so-called reaction-telegraph equation	
	\begin{align}
		\tau\frac{\partial^2 u}{\partial t^2}+\left( 1 - f^{\prime}(u) \right)\frac{\partial u}{\partial t}= d \nabla^{2}u+ f(u).
	\end{align}
	It is straightforward to generalize this derivation to two-species $u,v$. Doing so gives the coupled reaction-telegraph system: 
	\begin{equation}\begin{aligned}
			\tau_{1}\frac{\partial^2 u}{\partial t^2} + \frac{\partial u}{\partial t} - \tau_{1}\left(  \frac{\partial {f}(u,v)}{\partial u}\frac{\partial u}{\partial t} + \frac{\partial {f}(u,v)}{\partial v}\frac{\partial v}{\partial t} \right)
			= d_{11}\nabla^{2}u + {f}(u,v),
			\\
			\tau_{2}\frac{\partial^2 v}{\partial t^2} + \frac{\partial v}{\partial t} - \tau_{2}\left(  \frac{\partial {g}(u,v)}{\partial u}\frac{\partial u}{\partial t} + \frac{\partial {g}(u,v)}{\partial v}\frac{\partial v}{\partial t} \right) 
			= d_{22}\nabla^{2}v + {g}(u,v).
			\label{Eq:Reaction-Telegraph_Equations}	
	\end{aligned}\end{equation}
	We note that the nomenclature is not uniform in the literature, and sometimes hyperbolic reaction-diffusion systems, such as in Eq. \eqref{applicationgen}, are referred to as reaction-telegraph systems, and \eqref{Eq:Reaction-Telegraph_Equations} as reaction-Cattaneo systems. One major advantage to this model is that it can readily be derived from microscopic models of persistent random walks with reactions, as well as from a variety of thermodynamic formalisms such as generalized hydrodynamics \cite{al2004generalized} or extended irreversible thermodyanamics \cite{mendez1997dynamics}. We note that these systems do require sufficiently small relaxation times $\tau$ to maintain positivity of solutions, and for many chemical applications this is still a point of contention in the theory; see \cite{mendez2010reaction} for further discussion, as well as for further literature on studying this framework applied to chemical and ecological settings.

	Motivated by the aforementioned work, in the present paper we study a generic system of hyperbolic reaction-diffusion equations taking the form
	\begin{equation}\begin{aligned}\label{Eq:GeneralProblem}
			\tau_{1}\frac{\pa^2 u}{\pa t^2} & = \nabla \cdot \left( d_{11}\nabla u + d_{12}\nabla v \right) +f\left(\frac{\pa u}{\pa t},\frac{\pa v}{\pa t},u,v\right),\\
			\tau_{2}\frac{\pa^2 v}{\pa t^2} & = \nabla \cdot \left( d_{21}\nabla u + d_{22}\nabla v \right) +g\left(\frac{\pa u}{\pa t},\frac{\pa v}{\pa t},u,v\right),
	\end{aligned}\end{equation}
	where $\tau_{1},\tau_{2}\geq 0$ are constants, the $d_{11}=d_{11}(u,v)$ and $d_{22}=d_{22}(u,v)$ are self-diffusion parameters (or, Fickian diffusion when constant), the $d_{12}=d_{12}(u,v)$ and $d_{21}=d_{21}(u,v)$ are cross-diffusion parameters, and the functions $f$ and $g$ are assumed to have continuous derivatives in all arguments. For a complete picture of how all second derivatives influence diffusive instability leading to pattern formation, we have included self- and cross-diffusion parameters which in general depend on the unknown fields $u$ and $v$. Connections between pattern formation and either nonlinear diffusion or cross-diffusion were previously discussed in \cite{vanag2009cross, gambino2012turing, gambino2013turing, gambino2013pattern}, among many others. As pointed out earlier, diffusive instabilities in hyperbolic systems with cross-diffusion terms were recently studied in \cite{curro2021pattern}, and our model should account for such terms. We also allow for the appearance of the first time derivatives in a fairly generic manner. We consider the spatial domain to be a general smooth compact Riemannian manifold $\Omega$, with a suitable interpretation of the gradient, divergence, and Laplace-Beltrami operators (see \cite{krause2019influence, van2019turing, krause2021modern} for further details about reaction-diffusion systems posed on such manifolds). For cases where $\Omega$ has boundary, we take the no-flux boundary conditions $\mathbf{n}\cdot \left( d_{11}(u,v)\nabla u + d_{12}(u,v)\nabla v \right) = 0$ and $\mathbf{n} \cdot \left( d_{21}(u,v)\nabla u + d_{22}(u,v)\nabla v \right) =0$ on the boundary $\pa \Omega$, where $\mathbf{n}$ denotes the outward normal to the boundary.

	\section{Turing and wave instabilities}\label{sectheory}
	In this section, we outline the theory of diffusive instabilities for hyperbolic reaction-diffusion systems of the form \eqref{Eq:GeneralProblem}. We assume that there exist solutions $(u^*,v^*)$ of the algebraic system
	\be	\label{Eq:GeneralProblem_Homo}
	f(0,0,u^*,v^*)=0, 
	\quad
	g(0,0,u^*,v^*)=0,
	\ee
	which are spatially homogeneous, time constant solutions of \eqref{Eq:GeneralProblem}, $u=u^*$, $v=v^*$. A diffusive instability destabilizes such a spatially uniform solution, resulting in a new solution which is heterogeneous in space. There is a large and growing literature on pattern formation due to diffusive instabilities. In order to study the instability of such spatially uniform solutions, we consider perturbations 
	\be\label{perts}
	u = u^* +\epsilon U\,, \quad
	v = v^* +\epsilon V.
	\ee
	The corresponding linearised problem then reads
	\begin{align}\label{Eq:Linearizedproblem}
		T \frac{\pa^2}{\pa t^2}
		\left(
		\begin{array}{c}
			U  \\
			V
		\end{array}
		\right) + F \frac{\pa}{\pa t}
		\left(
		\begin{array}{c}
			U \\
			V
		\end{array}
		\right) = 
		D \nabla^2 \left(
		\begin{array}{c}
			U \\
			V
		\end{array}
		\right)
		+ J
		\left(
		\begin{array}{c}
			U  \\
			V
		\end{array}
		\right),
	\end{align}
	where we define the constant matrices 
	\begin{equation}\label{Eq:MatrixDef}
		\begin{aligned}
			T= \left( 
			\begin{array}{cc}
				\tau_{1} & 0 \\
				0 & \tau_{2}
			\end{array}
			\right), \quad D =
			\left( 
			\begin{array}{cc}
				d_{11} & d_{12} \\
				d_{21} & d_{22}
			\end{array}
			\right),\quad
			J = \begin{pmatrix}
				\dfrac{\partial f}{\partial {u}} & \dfrac{\partial f}{\partial {v}} \\
				\\
				\dfrac{\partial g}{\partial {u}} & \dfrac{\partial g}{\partial {v}}
			\end{pmatrix},\quad
			F = -\begin{pmatrix}
				\dfrac{\partial f}{\partial\dot{u}} & \dfrac{\partial f}{\partial \dot{v}} 	\\
				\\
				\dfrac{\partial g}{\partial \dot{u}} & \dfrac{\partial g}{\partial \dot{v}}
			\end{pmatrix} ,
		\end{aligned}
	\end{equation}
	with each entry evaluated at the homogeneous state $(0,0,u^*,v^*)$ and dots denoting time derivatives. 
	
	As \Eqref{Eq:Linearizedproblem} is linear, we decompose the perturbations $U$ and $V$ in terms of the eigenbasis for the underlying space domain. We denote the $\ell$th eigenfunction of $\nabla^2$ over $\Omega$ by $\Psi_{\ell}(\mathbf{x})$, where $\mathbf{x}$ is the collection of coordinates defined on $\Omega$. Note that $\ell$ can be a multi-index, if the dimension of the manifold $\Omega$ is greater than one, and this is a common occurrence for Turing analysis on manifolds of dimension greater than one \cite{van2019turing}. For a given choice of $\Omega$, there is a Neumann spectrum $\rho_{\ell}$ of the form $0=\rho_{0}<\rho_{1}<\rho_2 < \cdots \rightarrow \infty$ such that the eigenvalue problem $\nabla^2\Psi_{\ell}=-\rho_{\ell} \Psi_{\ell}$ holds. We choose the Neumann eigenbasis as this is appropriate for both manifolds without boundary and the no-flux conditions on manifolds with boundary. Writing the perturbations in terms of this eigenbasis, we have
	\begin{align}\label{pertseries}
		U=\sum_{\ell=0}^\infty U_{\ell} \mathrm{e}^{\lambda_\ell t} \Psi_{\ell}(\mathbf{x}),\;\;\; V=\sum_{\ell =0}^\infty V_{\ell} \mathrm{e}^{\lambda_\ell t} \Psi_{\ell}(\mathbf{x}),
	\end{align}
	where the $U_\ell$ and $V_\ell$ denote constants. Note that the $\ell =0$ terms (corresponding to $\rho_0 =0$) denote spatially uniform terms and may be omitted or treated separately, as their stability governs the linear stability of the spatially uniform base state to spatially uniform perturbations.
	
	Using the perturbation \eqref{pertseries} in \Eqref{Eq:Linearizedproblem}, we obtain for each eigenmode an algebraic problem
	\be \label{algmain}
	\left( \lambda_\ell^2 T + \lambda_\ell F + \rho_\ell D - J \right)\begin{pmatrix}	U_\ell  \\	V_\ell \end{pmatrix} = \begin{pmatrix}	0  \\	0 \end{pmatrix}.
	\ee
	The system \eqref{algmain} admits a non-trivial solution provided
	\be \label{detcondmain}
	\det\left(\lambda_\ell^2 T + \lambda_\ell F + \rho_\ell D - J\right) =0\,.
	\ee
	This is the condition by which we determine $\lambda_\ell$. As \eqref{detcondmain} is a polynomial equation in $\lambda_\ell$, understanding the structure of this polynomial will allow us to determine the sign of the real part of the roots $\lambda_\ell$, thereby providing information on whether the Turing instability occurs for a particular spatial mode $\Psi_{\ell}(\mathbf{x})$.
	
	Using \eqref{detcondmain} and expanding in $\lambda_\ell$, we obtain a degree four polynomial giving the characteristic equation for $\lambda_\ell$, which reads
	\begin{align} 
		& \det(T)\lambda_\ell^4 + (T,F)\lambda_\ell^3 + \left( (T,D)\rho_\ell + \det(F)-(T,J)\right)\lambda_\ell^2 \nonumber\\
		& +\left( (F,D)\rho_\ell - (F,J)\right)\lambda_\ell + \det(D)\rho_\ell^2 -(D,J)\rho_\ell +\det(J) =0\,.\label{poly1}
	\end{align}
	For ease of notation, we have made use of the product $(A, B)$ of $2\times 2$ matrices $A$ and $B$, which we define by
	\be 
	(A,B) = A_{11}B_{22} + A_{22}B_{11} - A_{12}B_{21} - A_{21}B_{12}\,.
	\ee 
	Clearly, $(A,B) = (B,A)$, so the product commutes. Observe that $(A,A)=2\det(A)$ and $(A,I) = \text{tr}(A)$, where $I$ is the $2\times 2$ identity matrix. As such, this product generalizes both the determinant and trace of a matrix.
	
	For compactness of notation, we define 
	\begin{subequations}\begin{align}
			\beta_0 &=\det(T),\\
			\beta_1 &=(T,F),\\
			\beta_2 &=(T,D)\rho_\ell + \det(F)-(T,J),\\
			\beta_3 &=(F,D)\rho_\ell - (F,J),\\
			\beta_4 &=\det(D)\rho_\ell^2 -(D,J)\rho_\ell +\det(J),
	\end{align}\end{subequations}
	so that \eqref{poly1} may be expressed as 
	\be \label{poly2}
	\beta_0 \lambda_\ell^4 + \beta_1 \lambda_\ell^3 + \beta_2 \lambda_\ell^2 + \beta_3 \lambda_\ell + \beta_4 =0\,.
	\ee
	Each $\beta_k$ depends on $\ell$ through $\rho_\ell$. By the Routh-Hurwitz stability criterion, we have that $\text{Re}(\lambda_\ell) <0$ for each of the four solutions $\lambda_\ell$ to \eqref{poly2} if and only if
	\begin{subequations}\label{b}\begin{align}
			\beta_1 & > 0\,,\label{b1}\\
			\beta_1\beta_2 - \beta_0\beta_3 & > 0\,,\label{b2}\\
			(\beta_1\beta_2 - \beta_0\beta_3)\beta_3 - \beta_1^2 \beta_4 & > 0\,,\label{b3}\\
			\beta_4 & > 0\label{b4}\,.
	\end{align}\end{subequations}
	If at least one of the conditions in \eqref{b} fails (e.g., has the reverse inequality), at least one of $\text{Re}(\lambda_\ell) > 0$. 
	
	In order to ensure that the spatially uniform base state $(u,v)=(u^*,v^*)$ is stable in time without diffusion, we require that each condition in \eqref{b} hold for $\ell =0$ (i.e., to $\rho_0 =0$). We then obtain  
	\begin{subequations}\label{sscond}\begin{align}
			(T,F)  > 0\,,\label{sscond1}\\
			(T,F)\left( \det(F) - (T,J)\right) +\det(T)(F,J)  > 0\,,\label{sscond2}\\
			-(F,J)\left((T,F)\left( \det(F) - (T,J)\right) +\det(T)(F,J)\right) - (T,F)^2 \det(J)  > 0\,,\label{sscond3}\\
			\det(J)  > 0\label{sscond4}\,.
	\end{align}\end{subequations}
	On the other hand, we desire unstable spatial perturbations occurring only for some finite collection of spectral parameters. If the instability is present for $\rho_\ell \rightarrow \infty$, then there may exist an unstable cascade leading to blow-up, rather than pattern formation. As such, we should also have stability in the limit where the spectral parameter tends to infinity. As \eqref{b1} is independent of the spectral parameter, we need only consider \eqref{b2} - \eqref{b4}, from which we obtain the conditions
	\begin{subequations}\label{infcond}\begin{align}
			(T,F)(T,D)-\det(T)(F,D) & > 0\,,\label{infcond2}\\
			\left((T,F)(T,D)-\det(T)(F,D)\right)(F,D) - (T,F)^2\det(D) & > 0\,,\label{infcond3}\\
			\det(D) & >0\,. \label{infcond4}
	\end{align}\end{subequations}
	
	Note that \eqref{sscond1} fixes \eqref{b1}, while \eqref{sscond2} and \eqref{infcond2} fix condition \eqref{b2}. As such, neither \eqref{b1} nor \eqref{b2} will change sign for any $0\leq \rho_\ell < \infty$. Conditions \eqref{b3} and \eqref{b4} are quadratic in $\rho_\ell$. The signs of the constant and quadratic terms in \eqref{b3} are fixed by \eqref{sscond3} and \eqref{infcond3}, respectively. Similarly, the signs of the constant and quadratic terms in \eqref{b4} are fixed by \eqref{sscond4} and \eqref{infcond4}, respectively. This leaves only the linear terms in both \eqref{b3} and \eqref{b4} as degrees of freedom.
	
	A necessary condition for the polynomial \eqref{poly2} to have four solutions with negative real part is that all coefficients are of the same sign (while \eqref{b} gives the sufficient conditions). From the linear form of $\beta_3 = (T,D)\rho_\ell + \det(F)-(T,J)$, we must have $\beta_3 >0$ for all $\ell$ in order for conditions \eqref{b} to hold both as $\ell =0$ and $\ell \rightarrow \infty$. In light of \eqref{b2} always holding, we have $(\beta_1\beta_2 - \beta_0\beta_3)\beta_3 >0$. This means that in order for condition \eqref{b3} to fail at some finite $\rho_\ell$, the only possibility is to have $\beta_4 >0$. This, however, forces \eqref{b4} to hold. As such, a spatial instability which violates \eqref{b3} is mutually exclusive from an instability which violates \eqref{b4}. We therefore classify these two instabilities separately. 
	
	\subsection{Turing instability conditions}\label{TI}
	Consider an instability which violates condition \eqref{b4} first. In terms of system parameters, \eqref{b4} reads $\det(D)\rho_\ell^2 -(D,J)\rho_\ell +\det(J) >0$, so an instability must reverse this inequality for some finite collection of $\rho_\ell$. Since $\det(D) >0$ by \eqref{infcond4} and $\det(J) >0$ by \eqref{sscond4}, a diffusive instability corresponding to $\beta_4 <0$ occurs only when 
	\be \label{TuringSC1}
	(D,J) > 2\sqrt{\det(D)\det(J)},
	\ee
	which follows from the discriminant of \eqref{b4}. For such a case, the spatially uniform state is unstable under a perturbation with the $\ell$th mode corresponding to $\rho_\ell$ provided that $0 < \rho_- < \rho_\ell < \rho_+ < \infty$, where
	\be \label{TuringSpec1}
	\rho_{\pm} = \frac{(D,J)\pm \sqrt{(D,J)^2 - 4\det(D)\det(J)}}{2\det(D)}\,.
	\ee
	
	Note that \eqref{TuringSC1} and \eqref{TuringSpec1} hold for all degenerate cases, as these are independent of the matrices $T$ and $F$. In the case where cross diffusion is absent ($d_{12}=d_{21}=0$), \eqref{TuringSC1} and \eqref{TuringSpec1} reduce respectively to
	\be
	d_{11}J_{22} + d_{22}J_{11} > 2\sqrt{d_{11}d_{22}\det(J)}
	\ee
	and
	\be \begin{aligned}
		\rho_{\pm} & = \frac{d_{11}J_{22} + d_{22}J_{11} }{2d_{11}d_{22}} \pm \frac{\sqrt{(d_{11}J_{22} + d_{22}J_{11})^2 - 4d_{11}d_{22}\det(J)}}{2d_{11}d_{22}}\,,
	\end{aligned}\ee
	which are exactly the sufficient conditions for the Turing instability in the classical setting. 
	
	Then, if $\beta_4 <0$ yet all other conditions in \eqref{b} hold, there is exactly one positive root by Descartes' rule of signs. Therefore, there is exactly one root $\lambda_\ell$ of \eqref{poly2} for which $\text{Re}(\lambda_\ell) >0$ when \eqref{TuringSC1} and \eqref{TuringSpec1} hold, and furthermore this root is a positive real number, signifying a stationary pattern. As such, parameter regimes satisfying \eqref{TuringSC1} and \eqref{TuringSpec1} result in a Turing instability. 
	
	\subsection{Wave instability conditions}\label{THI}
	We now consider the case where the condition \eqref{b3} is violated for some finite $\rho_\ell$, leading to a diffusive instability. As previously mentioned, this case is mutually exclusive from the Turing instability which occurs for $\beta_4 <0$, so we assume \eqref{b4} holds, giving $\beta_4 >0$. Furthermore, from \eqref{b1} we have $\beta_1 >0$, while $\beta_0 = \det(T) >0$. As mentioned previously, $\beta_3 >0$. Finally, \eqref{b2} requires $\beta_2 >0$. Each coefficient of the polynomial in \eqref{poly2} is positive, and without sign changes, Descartes' rule of signs implies that there are no positive roots to \eqref{poly2}. As such, the only possibility for an instability is to have a complex conjugate pair of $\lambda_\ell$ with $\text{Re}(\lambda_\ell) >0$.  
	
	In order to explore the bifurcation arising from \eqref{b3}, let us define a small parameter $\delta$ by 
	\be \label{delta1}
	(\beta_1\beta_2 - \beta_0\beta_3)\beta_3 - \beta_1^2 \beta_4  = -\delta \,.
	\ee
	If $\delta <0$, then \eqref{b3} holds, while if $\delta >0$ then \eqref{b3} fails and there is an instability. We are therefore interested in the bifurcation corresponding to $\delta =0$. As \eqref{b1} and \eqref{b2} must always hold, we rearrange \eqref{delta1} into the form
	\be \label{delta2}
	\beta_4 = \frac{(\beta_1\beta_2 - \beta_0\beta_3)\beta_3 + \delta}{\beta_1^2}\,.
	\ee
	When $\delta >0$, \eqref{b3} fails yet $\beta_4 >0$ and hence \eqref{b4} holds, consistent with the fact that the instabilities arising from \eqref{b3} and \eqref{b4} are mutually exclusive in parameter space. Replacing $\beta_4$ with the expression \eqref{delta2}, \eqref{poly2} becomes
	\be \label{poly3}
	\beta_0 \lambda_\ell^4 + \beta_1 \lambda_\ell^3 + \beta_2 \lambda_\ell^2 + \beta_3 \lambda_\ell + \frac{(\beta_1\beta_2 - \beta_0\beta_3)\beta_3 + \delta}{\beta_1^2} =0\,.
	\ee
	When $\delta =0$, we find that the four roots of \eqref{poly3} are
	\begin{subequations}\begin{align}
			\lambda_{\ell;1,2} & = \pm i\sqrt{\frac{\beta_3}{\beta_1}}\,,\label{root1}\\
			\lambda_{\ell;3,4} & = \frac{-\beta_1^2 \pm \sqrt{\beta_1^4 - 4\beta_0\beta_1\left( \beta_1\beta_2 - \beta_0\beta_3\right)}}{2\beta_0\beta_1}\label{root2}\,.
	\end{align}\end{subequations}
	Since \eqref{b2} holds, \eqref{root2} implies that $\text{Re}(\lambda_{\ell;3,4})<0$ for all $\ell$, given small $\delta$. Therefore, it is the pair of roots $\lambda_{\ell;1,2}$ in \eqref{root1} which are responsible for the bifurcation at $\delta =0$. Consider a perturbation development of these roots, so that 
	\be \label{rootpert}
	\lambda_{\ell;1,2} = \pm i\sqrt{\frac{\beta_3}{\beta_1}} + \eta_{1,2}\delta + \mathcal{O}\left( \delta^2\right)\,.
	\ee
	Placing \eqref{rootpert} into \eqref{poly3}, we find that the first order correction $\eta_{1,2}$ is given by
	\be
	\eta_{1,2} = \frac{1}{2}\frac{\beta_1\pm i(\beta_1\beta_3)^{-1/2}\left( \beta_1\beta_2 -2\beta_0\beta_3\right)}{\beta_1^3\beta_3 + \left( \beta_1\beta_2 -2\beta_0\beta_3\right)^2}\,.
	\ee
	This gives, at leading order, 
	\begin{subequations}
		\be \label{real}
		\text{Re}(\lambda_{\ell;1,2}) = \frac{\beta_1}{\beta_1^3\beta_3 + \left( \beta_1\beta_2 -2\beta_0\beta_3\right)^2}\frac{\delta}{2} + \mathcal{O}\left(\delta^2\right)\,,
		\ee
		\be \label{imag}
		\text{Im}(\lambda_{\ell;1,2}) = \sqrt{\frac{\beta_3}{\beta_1}} + \mathcal{O}\left(\delta\right)\,.
		\ee
	\end{subequations}
	The factor multiplying $\delta$ in \eqref{real} is positive. Therefore, as $\delta$ crosses zero from negative to positive, the real part of the temporal eigenvalues $\lambda_{\ell;1,2}$ tends from negative to positive, while the imaginary part of the temporal eigenvalues $\lambda_{\ell;1,2}$ remains positive. As such, an instability due to a change in the sign of condition \eqref{b3} develops due to a Hopf bifurcation \cite{hassard1981theory, marsden2012hopf}. 
	
	Having better understood the bifurcation resulting in a loss of stability due to \eqref{b3}, we now obtain sufficient conditions for this instability to occur under a spatial perturbation. To this end, we write 
	\be \label{THpoly}
	(\beta_1\beta_2 - \beta_0\beta_3)\beta_3 - \beta_1^2 \beta_4 = \gamma_0 \rho_\ell^2 + \gamma_1 \rho_\ell + \gamma_2\,,
	\ee
	where
	\begin{subequations}\begin{align}
			\gamma_0  = & \left((T,F)(T,D)-\det(T)(F,D)\right)(F,D) - (T,F)^2\det(D)  > 0\,,\\
			\gamma_1  = & \left\lbrace (T,F)\left(\det(F)-(T,J)\right)+\det(T)(F,J) \right\rbrace (F,D)  \\
			&- \left\lbrace (T,F)(T,D)-\det(T)(F,D) \right\rbrace (F,J) + (T,F)^2 (D,J)\,,\label{gamma1}
			\nonumber
			\\ 
			\gamma_2  = & -(F,J)\left((T,F)\left( \det(F) - (T,J)\right) +\det(T)(F,J)\right) - (T,F)^2 \det(J)  > 0\,,
	\end{align}\end{subequations}
	are combinations of the system parameters. From \eqref{infcond3} and \eqref{sscond3} we have that $\gamma_0$ and $\gamma_2$ are positive, respectively. In order for \eqref{THpoly} to be negative for some $\rho_\ell$, a necessary condition is that $\gamma_1 <0$. The first two terms in \eqref{gamma1} are positive (as can be shown using \eqref{infcond} and \eqref{sscond}). Therefore, a necessary condition for $\gamma_1 <0$ is that $(D,J)<0$. This is in contrast to the Turing case, where we found \eqref{TuringSC1} and hence that $(D,J)>0$. This is consistent with the fact that the two instabilities are mutually exclusive.
	
	The discriminant of \eqref{b3} gives the following sufficient condition for the wave instability:
	\be \label{THcond1}
	\gamma_1 < -2\sqrt{\gamma_0 \gamma_2}<0,
	\ee
	which again is only possible provided $(D,J)<0$. For such a case, the spatially uniform state is unstable under a perturbation with the $\ell$th mode corresponding to $\rho_\ell$ provided that $0 < \hat{\rho}_- < \rho_\ell < \hat{\rho}_+ < \infty$, where
	\be \label{THcond2}
	\hat{\rho}_\pm = \frac{-\gamma_1 \pm \sqrt{\gamma_1^2 - 4\gamma_0\gamma_2}}{2\gamma_0}\,.
	\ee
	
	To gain an intuition for how the conditions for the wave instability differ from those for the Turing instability, consider the case without cross-diffusion, so that $D = \text{diag}(d_{11},d_{22})$. Since $\text{tr}(J)<0$, we assume without loss of generality that $J_{11}>0$, $J_{22}<0$, with $J_{11} < |J_{22}|$, so that $u$ is the activator and $v$ is the inhibitor. Then, the Turing condition $(D,J)>0$ is equivalent to $d_{22}>(|J_{22}|/J_{11})d_{11}$, hence the inhibitor tends to diffuse much faster than the activator (assuming the entries of the Jacobian matrix are $\mathcal{O}(1)$). On the other hand, the wave instability condition $(D,J)<0$ implies that $d_{22}<(|J_{22}|/J_{11})d_{11}$, which allows for scenarios where the activator diffuses faster than the inhibitor.
	
	\subsection{The zero-friction limit}
	Having considered the general hyperbolic system, it is now instructive to consider several physically relevant degenerate limits. We begin with the zero-friction limit, which is represented by a model of the form
	\begin{subequations}\label{zerofriction}
		\begin{align}
			\tau_{1}\frac{\pa^2 u}{\pa t^2} & = \nabla \cdot \left( d_{11}(u,v)\nabla u + d_{12}(u,v)\nabla v \right)+f\left(u,v\right),\\
			\tau_{2}\frac{\pa^2 v}{\pa t^2} & = \nabla \cdot \left( d_{21}(u,v)\nabla u + d_{22}(u,v)\nabla v \right)+g\left(u,v\right),
		\end{align}
	\end{subequations}
	In the zero-friction limit, we set $F=0_{2\times 2}$, the $2\times 2$ matrix with zero entries. The characteristic polynomial \eqref{poly1} reduces to 
	\be\begin{aligned} \label{polydeg1}
		\det(T)\lambda_\ell^4 + \left( (T,D)\rho_\ell -(T,J)\right)\lambda_\ell^2 + \det(D)\rho_\ell^2 -(D,J)\rho_\ell +\det(J) =0\,.
	\end{aligned}\ee
	At the base state, $\rho=0$, and this polynomial reduces to 
	\be\begin{aligned} \label{polydeg1a}
		& \det(T)\lambda_\ell^4  - (T,J)\lambda_\ell^2 + \det(J) =0\,.
	\end{aligned}\ee
	Such a polynomial is never stable; at best, there exist complex roots with zero real part. This is because the stability condition \eqref{sscond3} is violated. As such, a spatial perturbation does not result in a Turing or wave type bifurcation, as the base state itself is already unstable. This was also pointed out for the case of $\tau_1 = \tau_2$ in \cite{zemskov2016diffusive}.
	
	For such a case, the natural base state would no longer be a stable spatially uniform steady state, but rather a spatially uniform stable limit cycle. Although one could make progress in the case of a simple limit cycle arising from a linear system, the proper base state will arise as the solution of the coupled nonlinear ordinary differential equations found when neglecting diffusion terms, rather than a constant base state $u=u^*$, $v=v^*$. The nonlinearity prohibits the existence of a simple limit cycle which scales like $e^{i\omega t}$. As the base case in such a situation varies in time and is governed by a nonlinear process, a more sophisticated approach would be required to consider spatial perturbations about the stable limit cycle. This is a topic of ongoing work.

	\subsection{Hyperbolic-parabolic limit}
	It is possible to consider the case where only one of the reaction-diffusion equations is hyperbolic, with the other taking the standard parabolic form. Such a system can be written as
	\begin{subequations}\label{hp}
		\begin{align}
			\tau_{1}\frac{\pa^2 u}{\pa t^2} & = \nabla \cdot \left( d_{11}(u,v)\nabla u + d_{12}(u,v)\nabla v \right) +f\left(\frac{\pa u}{\pa t},\frac{\pa v}{\pa t},u,v\right),\\
			0 & = \nabla \cdot \left( d_{21}(u,v)\nabla u + d_{22}(u,v)\nabla v \right) +g\left(\frac{\pa u}{\pa t},\frac{\pa v}{\pa t},u,v\right)\,.
		\end{align}
	\end{subequations}
	To explore the limit where one of the equations in system \eqref{Eq:GeneralProblem} is parabolic, we set $\tau_2=0$. For this case, the characteristic polynomial \eqref{poly1} becomes
	\be \label{polydeg2}
	\alpha_0 \lambda_\ell^3 + \alpha_1\lambda_\ell^2 +\alpha_2\lambda_\ell + \alpha_3 =0\,,
	\ee
	where 
	\begin{subequations}\begin{align}
			\alpha_0 &= \tau_1 F_{22}\,,\\
			\alpha_1 &= \tau_1 d_{22}\rho_\ell + \det(F)-\tau_1 J_{22}\,,\\
			\alpha_2 &= (F,D)\rho_\ell - (F,J)\,,\\
			\alpha_3 &=\det(D)\rho_\ell^2 -(D,J)\rho_\ell +\det(J).
	\end{align}\end{subequations}
	We assume $F_{22} >0$ so that $\alpha_0 >0$. In fact, provided the diffusion coefficients have a fixed sign, it is possible to show that \eqref{hp} is a well-posed parabolic-hyperbolic system only if $F_{22},
	\tau_{1}>0$. We therefore restrict to this case without loss of generality. By the Routh-Hurwitz stability criterion, the polynomial \eqref{polydeg2} is stable if and only if
	\begin{subequations}\label{hpcond1}\begin{align}
			\tau_1 d_{22}\rho_\ell + \det(F)-\tau_1 J_{22}&>0\,,
			\label{hpc1a}
			\\
			\left( \tau_1 d_{22}\rho_\ell + \det(F)-\tau_1 J_{22}\right)\left( (F,D)\rho_\ell - (F,J)\right)&
			\label{hpc1b}
			\\
			-\tau_1 F_{22}\left(\det(D)\rho_\ell^2 -(D,J)\rho_\ell +\det(J)\right)&>0\,,
			\nonumber
			\\
			\det(D)\rho_\ell^2 -(D,J)\rho_\ell +\det(J)&>0\,.
			\label{hpc1c}
	\end{align}\end{subequations}
	In order for the base state to be stable, \eqref{hpcond1} must be stable at $\rho_0=0$, which gives the set of conditions
	\begin{subequations}\label{hpcond2}\begin{align}
			\det(F)-\tau_1 J_{22}>0\,,\label{hpc2a}\\
			-\left(\det(F)-\tau_1 J_{22}\right)(F,J)-\tau_1 F_{22}\det(J)>0\,,\label{hpc2b}\\
			\det(J)>0\,.\label{hpc2c}
	\end{align}\end{subequations}
	In order to have stable dynamics as $\rho_\ell \rightarrow \infty$, we must have the additional conditions
	\begin{subequations}\label{hpcond3}\begin{align}
			\tau_1 d_{22}>0\,,\label{hpc3a}\\
			\tau_1 d_{22} (F,D)-\tau_1 F_{22}\det(D)>0\,,\label{hpc3b}\\
			\det(D)>0\,.\label{hpc3c}
	\end{align}\end{subequations}
	Note that \eqref{hpc2a} and \eqref{hpc3a} imply that \eqref{hpc1a} always holds, so the only possibly instabilities arise from failure of either \eqref{hpc1b} or \eqref{hpc1c}. In particular we have that $\alpha_{1},\alpha_{2}>0$
	
	The instability resulting in failure of \eqref{hpc1c} is exactly the Turing instability classified in Sec. \ref{TI}, with the parameter condition being \eqref{TuringSC1}, and the range of spectral parameters $\rho_\ell$ resulting in the Turing instability given by \eqref{TuringSpec1}, with the only change being the modified conditions for stability of the base state \eqref{hpcond2} and stability for asymptotically large spectra \eqref{hpcond3}.
	
	In order to explore the nature of an instability arising from \eqref{hpc1b}, let us define the small parameter $\delta$ through the relation
	\be 
	\alpha_1\alpha_2 - \alpha_0\alpha_3 = -\delta\,.
	\ee
	Isolating for $\alpha_3$ and placing the resulting expression into \eqref{polydeg2}, we find
	\be \label{deltaalpha}
	\alpha_0 \lambda_\ell^3 + \alpha_1 \lambda_\ell^2 + \alpha_2 \lambda_\ell + \frac{\alpha_1\alpha_2 +\delta}{\alpha_0} =0\,.
	\ee
	Taking $\delta=0$, we find the roots 
	\be 
	\lambda_{\ell; 1,2} = \pm i \sqrt{\frac{\alpha_2}{\alpha_0}}\,,\quad \lambda_{\ell;3} = - \frac{\alpha_1}{\alpha_0}\,.
	\ee
	The roots $\lambda_{\ell;1,2}$ are purely imaginary, while the third root is always negative. Considering a perturbation expansion 
	\be 
	\lambda_{\ell; 1,2} = \pm i \sqrt{\frac{\alpha_2}{\alpha_0}}+ \hat{\eta}_{\pm}\delta + \mathcal{O}\left( \delta^2\right)\,,
	\ee
	in \eqref{deltaalpha}, we find 
	\be 
	\hat{\eta}_\pm = \frac{1}{2\left(\alpha_0\alpha_2+\alpha_1^2\right)}+\frac{i \alpha_1}{2\sqrt{\alpha_0\alpha_2}\left(\alpha_0\alpha_2+\alpha_1^2\right)}\,.
	\ee
	Therefore, at leading order,
	\be 
	\text{Re}(\lambda_{\ell; 1,2}) = \frac{1}{\left(\alpha_0\alpha_2+\alpha_1^2\right)}\frac{\delta}{2} + \mathcal{O}\left( \delta^2\right)\,,
	\ee
	\be 
	\text{Im}(\lambda_{\ell; 1,2}) = \pm i \sqrt{\frac{\alpha_2}{\alpha_0}} + \mathcal{O}\left( \delta\right)\,,
	\ee
	so near the bifurcation $\delta =0$ the imaginary part of the temporal eigenvalues $\lambda_{\ell; 1,2}$ remains non-zero, while the real part transitions from negative to positive (since $\alpha_0\alpha_2+\alpha_1^{2}>0$) as $\delta$ increases. This corresponds to a bifurcation leading to a wave instability, akin to that possible for the full hyperbolic-hyperbolic system which was classified in Sec. \ref{THI}. 
	
	In order to obtain the parameter set giving sufficient conditions for this instability, we write 
	\be \label{hpTHpoly}
	\alpha_1\alpha_2 - \alpha_0\alpha_3 = \hat{\gamma}_0 \rho_\ell^2 + \hat{\gamma_1} \rho_\ell + \hat{\gamma}_2\,,
	\ee
	where
	\begin{subequations}\begin{align}
			\hat{\gamma}_0 & = \tau_1 \left( d_{22} (F,D)- F_{22}\det(D)\right)>0\,,\\
			\hat{\gamma}_1 & = -\tau_1d_{22}(F,J)+\left(\det(F)-\tau_1J_{22}\right)(F,D) +\tau_1F_{22}(D,J)\,,\label{hpgamma1}\\ 
			\hat{\gamma}_2 & = -\left(\det(F)-\tau_1 J_{22}\right)(F,J)-\tau_1 F_{22}\det(J)>0\,,
	\end{align}\end{subequations}
	are combinations of the system parameters. From \eqref{hpc2b} and \eqref{hpc3b} we have that $\hat{\gamma}_0$ and $\hat{\gamma}_2$ are positive. In order for \eqref{hpTHpoly} to be negative for some $\rho_\ell$, a necessary condition is then that $\hat{\gamma}_1 <0$. The first two terms in \eqref{hpgamma1} are positive (as can be shown using \eqref{hpcond2} and \eqref{hpcond3}). Therefore, a necessary condition for $\hat{\gamma}_1 <0$ is that $(D,J)<0$. 
	
	The sufficient condition for the wave instability, which is found from the discriminant of \eqref{hpTHpoly}, is
	\be \label{hpTHcond1}
	\hat{\gamma}_1 < -2\sqrt{\hat{\gamma}_0 \hat{\gamma}_2},
	\ee
	which again is only possible provided $(D,J)<0$. Assuming \eqref{hpTHcond1} holds, the spatially uniform state is then unstable under a perturbation with the $\ell$th mode corresponding to $\rho_\ell$ provided that $0 < \tilde{\rho}_- < \rho_\ell < \tilde{\rho}_+ < \infty$, where
	\be \label{hpTHcond2}
	\tilde{\rho}_\pm = \frac{-\hat{\gamma}_1 \pm \sqrt{\hat{\gamma}_1^2 - 4\hat{\gamma}_0\hat{\gamma}_2}}{2\hat{\gamma}_0}\,.
	\ee
	Structurally, \eqref{hpTHcond1}-\eqref{hpTHcond2} are the same as conditions \eqref{THcond1}-\eqref{THcond2} for the full hyperbolic-hyperbolic system. The difference here is in the modified parameter sets resulting from the degeneracy $\tau_2 =0$.
	
	The reduction to a hyperbolic-parabolic system preserves the existence of both the Turing and wave type instabilities seen in the full hyperbolic-hyperbolic system. The quantitative differences arise from taking $\tau_2 =0$, which reduces the number of conditions needed to ensure stability of both the base state and of states found as $\rho_\ell \rightarrow \infty$, and also changes the parameter set for the wave bifurcation.
	
	\subsection{Parabolic-parabolic limit}
	The prototypical model for the Turing instability is a parabolic-parabolic reaction-diffusion system. Taking $\tau_1 = \tau_2 =0$, we recover from \eqref{Eq:GeneralProblem} a generalized (in the sense that both equations may involve both time derivatives) parabolic-parabolic system of the form
	\begin{subequations}\label{pp}
		\begin{align}
			\nabla \cdot \left( d_{11}(u,v)\nabla u + d_{12}(u,v)\nabla v \right)+f\left(\frac{\pa u}{\pa t},\frac{\pa v}{\pa t},u,v\right)&=0,\\
			\nabla \cdot \left( d_{21}(u,v)\nabla u + d_{22}(u,v)\nabla v \right)+g\left(\frac{\pa u}{\pa t},\frac{\pa v}{\pa t},u,v\right)&=0.
		\end{align}
	\end{subequations}
	In this limit, we set $\tau_1 = \tau_2 =0$, giving $\det(T)=0$ and $(T,A)=0$ for any $2\times 2$ matrix $A$. We obtain the characteristic polynomial
	\be\begin{aligned} \label{polydeg3}
		& \det(F)\lambda_\ell^2 +\left( (F,D)\rho_\ell - (F,J)\right)\lambda_\ell + \det(D)\rho_\ell^2 -(D,J)\rho_\ell +\det(J) =0\,.
	\end{aligned}\ee
	The Routh-Hurwitz stability criterion tells us that the polynomial \eqref{polydeg3} is stable if and only if all coefficients are positive. The base state (corresponding to $\rho_0 =0$) is then stable provided $\det(F)>0$, $(F,J)<0$, and $\det(J)>0$. For stability at asymptotically large $\rho_\ell$, we additionally require $(F,D) >0$ and $\det(D)>0$. In the classical case of two reaction-diffusion systems which are linear in the first-order time derivatives, we may write $F=I$, the $2\times 2$ identity matrix, and these conditions reduce to $\text{tr}(J) <0$, $\det(J)>0$, $\text{tr}(D)>0$, $\det(D) >0$.
	
	From the above conditions, the quadratic and linear terms in \eqref{polydeg3} are fixed, and only the constant term can change sign. Such a sign change can only occur if $(D,J) >0$, and a sufficient condition is exactly the same as was found before in \eqref{TuringSC1}. Similarly, the range of spectral parameters $\rho_\ell$ resulting in the Turing instability is exactly the same as in \eqref{TuringSpec1}. The only change in the parabolic-parabolic limit is then the modified set of conditions granting the stability of the polynomial \eqref{polydeg3} in the limits $\rho_\ell =0$ (stable base state) and $\rho_\ell \rightarrow \infty$. 
	
	The second class of instability, corresponding to \eqref{THcond1}-\eqref{THcond2}, does not occur for the parabolic-parabolic reduction. 
	
	\subsection{Hyperbolic-elliptic limit}
	It is possible to consider the limit where the $v$ species is governed by an elliptic equation. The proper way to write such a system is 
	\begin{subequations}\label{he1}
		\begin{align}
			\tau_{1}\frac{\pa^2 u}{\pa t^2} & = \nabla \cdot \left( d_{11}(u,v)\nabla u + d_{12}(u,v)\nabla v \right) +f\left(\frac{\pa u}{\pa t},u,v\right),\\
			0 & = \nabla \cdot \left( d_{21}(u,v)\nabla u + d_{22}(u,v)\nabla v \right) +g\left(u,v\right).
		\end{align}
	\end{subequations}
	Our prior linear perturbation analysis is still applicable if we set $\tau_2=0$, $F_{12}=F_{21}=F_{22}=0$. In this case, \eqref{poly1} reduces to
	\be\begin{aligned} \label{he2}
		& \tau_1\left(d_{22}\rho_\ell -J_{22} \right) \lambda_\ell^2 + F_{11}\left(d_{22}\rho_\ell -J_{22} \right)\lambda_\ell  +\det(D)\rho_\ell^2 -(D,J)\rho_\ell +\det(J)=0\,.
	\end{aligned}\ee
	The polynomial \eqref{he2} is stable at the base state ($\rho_0=0$) and for asymptotically large spectrum ($\rho_\ell \rightarrow \infty$) provided 
	\begin{align}
		\tau_1 >0,
		\quad
		F_{11}>0, 
		\quad
		J_{22} <0,
		\quad 
		\det(D)>0,
		\quad 
		\det(J) >0.
	\end{align}

	The only possibility for instability is if the term $\det(D)\rho_\ell^2 -(D,J)\rho_\ell +\det(J)$ changes sign from positive to negative, meaning that a necessary condition is $(D,J)>0$. This is exactly the same as the Turing instability described in Sec. \ref{TI}, and the sufficient conditions remain \eqref{TuringSC1}-\eqref{TuringSpec1}. The only difference is that the parameter conditions arising from requiring stability of the base state and the asymptotically large spectrum limit are modified.
	
	In the standard Turing instability for parabolic-parabolic systems, one of $J_{11}$ or $J_{22}$ must be negative, and the other positive. In the hyperbolic-parabolic system, it must be the $J_{11}$ from the hyperbolic species which is positive, and the $J_{22}$ from the elliptic species which is negative. Therefore, the hyperbolic-elliptic system will admit diffusive instability under a more restricted set of reaction kinetics, compared to the parabolic-parabolic counterpart.

	\subsection{Scalar hyperbolic limit}
	As we have shown in considering several reductions of the full hyperbolic-hyperbolic model, diffusive instabilities may occur when the characteristic polynomial \eqref{poly1} is of degree two or greater, with the degree two case giving patterns in the classical parabolic-parabolic case, as well as the hyperbolic-elliptic case. It is natural to wonder if diffusive instabilities then result from a scalar hyperbolic model. The scalar hyperbolic reduction of \eqref{Eq:GeneralProblem} will take the form
	\be \label{scalar1}
	\tau\frac{\pa^2 u}{\pa t^2} = \nabla \cdot \left( d(u)\nabla u\right) +f\left(\frac{\pa u}{\pa t},u\right)\,,
	\ee
	where $\tau >0$ and $f$ is differentiable in both variables. Linearizing about the base state $u =u^*$ satisfying $f(0,u^*)=0$, we obtain the characteristic polynomial
	\be \label{scalar2}
	\tau \lambda_\ell^2 + F \lambda_\ell + d(u^*)\rho_\ell - J =0\,,
	\ee
	where we define the constants $F = -\frac{\pa f}{\pa \dot{u}}$ and $J = \frac{\pa f}{\pa u}$ with evaluation at the base state $u=u^*$. In order for the polynomial \eqref{scalar2} to be stable at the base state (corresponding to $\rho_0 =0$), we require $F >0$ and $J<0$. Yet, we then have that $d(u^*)\rho_\ell -J \geq -J >0$ for all $\ell$, and as such \eqref{scalar2} never loses stability for any $\ell$. 
	
	This case shows that having a scalar equation with higher-order derivatives in time is not sufficient for diffusive instabilities. Although at least two time derivatives are required, an addition requirement is that there is some competition between spatial modes in different species, which mathematically result in a quadratic polynomial in the spectral parameter $\rho_\ell$. Without this, a diffusive instability cannot occur.

	\section{Pattern formation under the Turing and wave instabilities}\label{secexamples}
	We now present numerical simulations of systems of the form \eqref{Eq:GeneralProblem} to better understand how the diffusive instabilities outlined in Sec. \ref{sectheory} influence the fully nonlinear dynamics of hyperbolic reaction-diffusion systems. 
	
	\subsection{Turing patterns under the Turing instability}\label{secturingappl}
	In order to demonstrate how patterns eventually evolve from diffusive instabilities, we consider concrete examples of reaction-diffusion processes. In the present section, we focus on the Turing instability, as described in Sec. \ref{TI}. To illustrate the theory developed in Sec. \ref{TI}, we consider the generic hyperbolic reaction-diffusion system
	\begin{subequations}\label{applicationgen}
		\begin{align}
			\tau_{1}\frac{\pa^2 u}{\pa t^2} + F_{11}\frac{\pa u}{\pa t} + F_{12}\frac{\pa v}{\pa t} & = d_{11}\nabla^2 u  +f\left(u,v\right),\\
			\tau_{2}\frac{\pa^2 v}{\pa t^2} + F_{21}\frac{\pa u}{\pa t} + F_{22}\frac{\pa v}{\pa t} & = d_{22}\nabla^2 v +g\left(u,v\right).
		\end{align}
	\end{subequations}
	Here the $\tau_{k}$, $F_{k\ell}$, and $d_{k\ell}$ are constant parameters, while $f$ and $g$ are assumed to be continuously differentiable in their arguments. We choose $d_{11}$, $d_{22}$, $f$, and $g$ from known pattern forming systems, while the other parameters will be treated as control parameters.
	
	First consider the standard Turing system
	\begin{subequations}\label{standard}
		\begin{align}
			\frac{\pa u}{\pa t} & = d_{11}\nabla^2 u +f\left(u,v\right),\\
			\frac{\pa v}{\pa t} & = d_{22}\nabla^2 v +g\left(u,v\right).
		\end{align}
	\end{subequations}
	We employ three sets of reaction kinetics and diffusion parameters. \\
	\\
	\noindent I. Schnakenberg kinetics and diffusion rates
	\begin{subequations}\label{schparams}\be 
		f(u,v) = \frac{1}{10}-u+u^2v\,, \quad g(u,v) = 1-u^2v\,,
		\ee
		\be 
		d_{11}=0.0005\,, \quad d_{22} = 0.01\,.
		\ee\end{subequations} 
	A unique locally stable steady state is given by $u^* = 11/10$, $v^* = 100/121$. When the domain permits unstable wavenumbers, a Turing instability of this steady state occurs, resulting in spot patterns occurring for $u$.\\
	\\
	\noindent II. FitzHugh-Nagumo kinetics and diffusion rates
	\begin{subequations}\label{fnparams}\be 
		f(u,v) = u-0.33u^3+v-0.6\,, \quad g(u,v) = 0.6-u-0.99v\,,
		\ee
		\be 
		d_{11}=0.00008\,, \quad d_{22} = 0.004\,.
		\ee\end{subequations}
	A unique locally stable steady state is given by $u^* = 0.22548$, $v^* = 0.37830$. When the domain permits unstable wavenumbers, a Turing instability of this steady state occurs, leading to the formation of stripe patterns in $u$. We note that the $u$ variable in these kinetics often represents a difference from a fixed voltage potential \cite{keener1998mathematical}, so we do not require strictly non-negative concentrations as one often does in chemical or biological applications.\\
	\\
	\noindent III. The kinetics and diffusion rates
	\begin{subequations}\label{sunkenparams}\be 
		f(u,v) = 0.75vu^3-0.85u^3+0.85u^2-1.65u\,, 
		\ee
		\be 
		g(u,v) = -0.75u^2v^2+0.75u^2v-0.75uv+2.25v\,,
		\ee
		\be 
		d_{11}=0.0005\,, \quad d_{22} = 0.05\,,
		\ee\end{subequations}
	were studied in \cite{van2020turing}. A locally stable steady state is given by $u^* = 3$, $v^* = 1$. When the domain permits unstable wavenumbers, a Turing instability of this steady state occurs, giving rise to sunken spots in the field $u$. 
	
	\begin{figure}[t!]
		\begin{center}
			\includegraphics[width=0.9\textwidth]{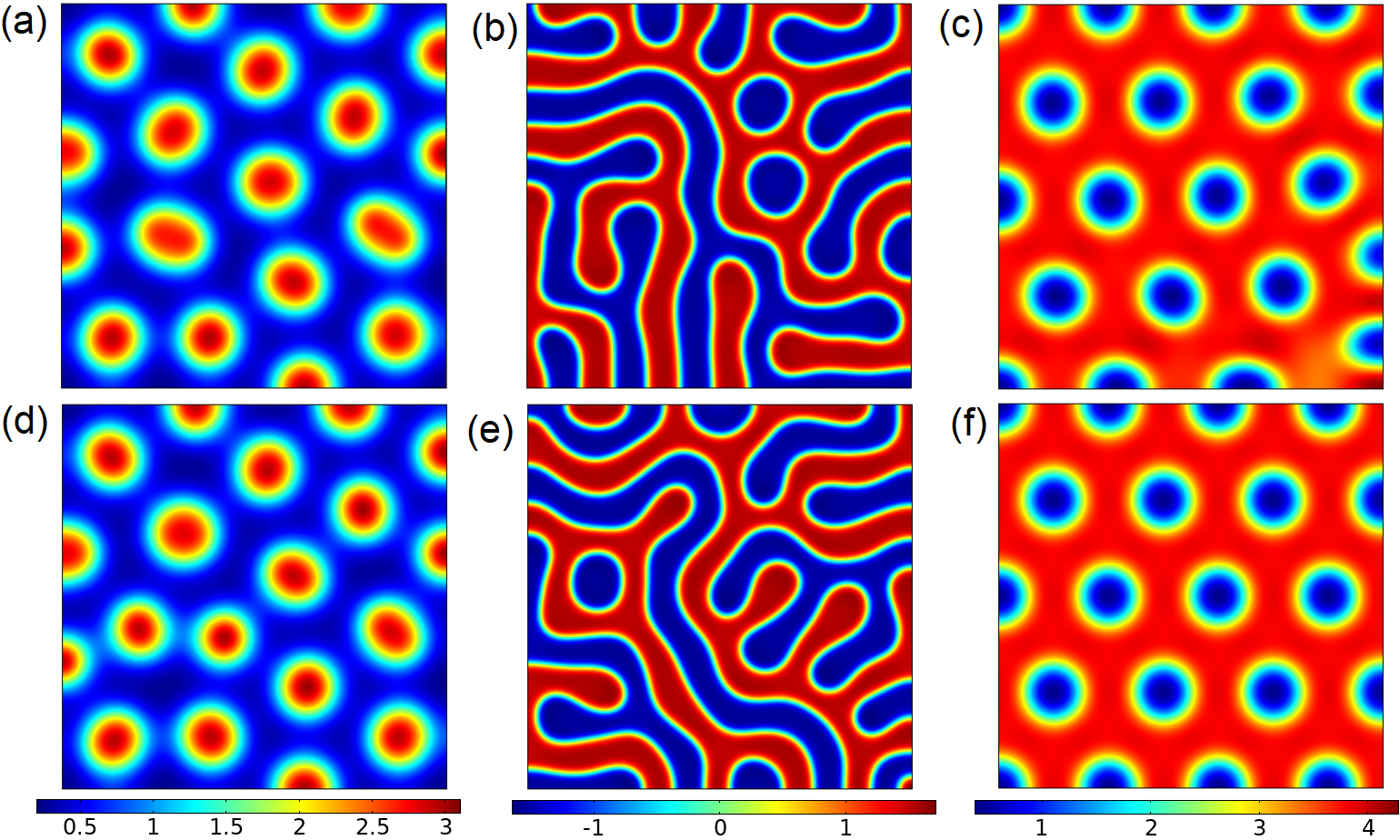}
			\vspace*{-0.15in}
			\caption{Turing patterns corresponding to the function $u$ at $t=10000$ emergent from (a,d) Schnakenberg kinetics $f(u,v) = 0.1-u+u^2v$, $g(u,v) = 1-u^2v$, $d_{11}=0.0005$, $d_{22} = 0.01$ giving spot patterns; (b,e) FitzHugh-Nagumo kinetics $f(u,v) = u-0.33u^3+v-0.6$, $g(u,v) = 0.6-u-0.99v$, $d_{11}=0.00008$, $d_{22} = 0.004$ giving stripe patterns (the $u$ variable in these kinetics often represents a difference from a fixed voltage potential \cite{keener1998mathematical}, so we do not require strictly non-negative concentrations as one often does in chemical or biological applications); (c,f) the kinetics  $f(u,v) = 0.75vu^3-0.85u^3+0.85u^2-1.65u$, $g(u,v) = -0.75u^2v^2+0.75u^2v-0.75uv+2.25v$, $d_{11}=0.0005$, $d_{22} = 0.05$ giving sunken spots \cite{van2020turing}. In (a)-(c) we consider the standard Turing system with $\tau_1=\tau_2=F_{12}=F_{21}=0$ taken to generate each plot. Then, for each respective set of reaction kinetics, we vary parameters like (d) $\tau_1=\tau_2=0.1$, $F_{12}=F_{21}=0.5$, (e) $\tau_1=\tau_2=0.2$, $F_{12}=F_{21}=0.5$, (f) $\tau_1=\tau_2=0.2$, $F_{12}=F_{21}=1$. In all cases we take $F_{11}=F_{22}=1$.	 Each column uses the color scale shown at the bottom. Although the temporal parameters control the size of the Turing space, once a pattern actually forms these parameters do not play a strong role in the final steady state pattern. \label{Fig1}}
		\end{center}
	\end{figure}

	We simulate the hyperbolic reaction-diffusion system \eqref{applicationgen} under three different sets of reaction kinetics, as shown in Fig. \ref{Fig1}. To generate these numerical solutions, we solved \eqref{applicationgen} subject to no-flux boundary conditions on the unit square $[0,1]^2$. The initial data was taken to be of the form $u=u^*\zeta(\mathbf{x})$, $v = v^*\xi(\mathbf{x})$ where $u^*,v^*$ are the homogeneous steady states of the kinetics, and $\zeta$ and $\xi$ are normally distributed random variables at each spatial point with mean $1$ and standard deviation $10^{-2}$. The equations were solved using the finite-element software COMSOL, with 50786 second-order triangular elements (as will be the case for all subsequent simulations on square domains). We used a backwards-differentiation formula (BDF) of orders 1-5 as the timestepping scheme, with a tolerance of $10^{-4}$. Solutions to classical models with all $\tau_k$, $F_{k\ell}$ equal to zero are shown in Fig. \ref{Fig1}(a-c). Using the same initial data (i.e., the same random realization of the perturbations), we explored a variety of both the temporal parameters, $\tau_k$ and $F_{k\ell}$. Examples of patterns found inside of these regions are shown in Fig. \ref{Fig1}(d-f). Although they restrict the region permitting the Turing instability, we find that the $\tau_k$ and $F_{k\ell}$ parameters have little effect on the qualitative structure of stationary patterns observed. Differences between these simulations are qualitatively the same as using different random initial perturbations, for which such systems are known to be very sensitive.

	\subsection{Turing patterns in reaction-telegraph systems}\label{secreacttele}
	In order to determine the nature of diffusive instabilities from reaction-telegraph systems \eqref{Eq:Reaction-Telegraph_Equations}, we again apply the theory of Sec. \ref{sectheory}. The corresponding linearisation is of the form \eqref{Eq:Linearizedproblem}, with $D=\text{diag}(d_{11},d_{22})$ and 
	\be
	F = \left(
	\begin{array}{cc}
		1-\tau_{1}J_{11} & -\tau_{1}J_{12}
		\\
		-\tau_{2}J_{21} & 1-\tau_{2}J_{22} 
	\end{array}	
	\right).
	\ee
	The set of conditions \eqref{TuringSC1}-\eqref{TuringSpec1} remain unchanged, and so the set of unstable modes leading to the Turing instability is independent of $T$ and $F$. As in Sec. \ref{secturingappl}, the only way for the $T$ or $F$ matrices to influence Turing pattern formation is through the feasibility conditions. Unlike in Sec. \ref{secturingappl}, the entries in $F$ are completely determined by the entries of $J$ and $T$. As such, the only remaining control parameters are $\tau_1$ and $\tau_2$. 
	
	\begin{figure}[t!]
		\begin{center}
			\includegraphics[width=0.9\textwidth]{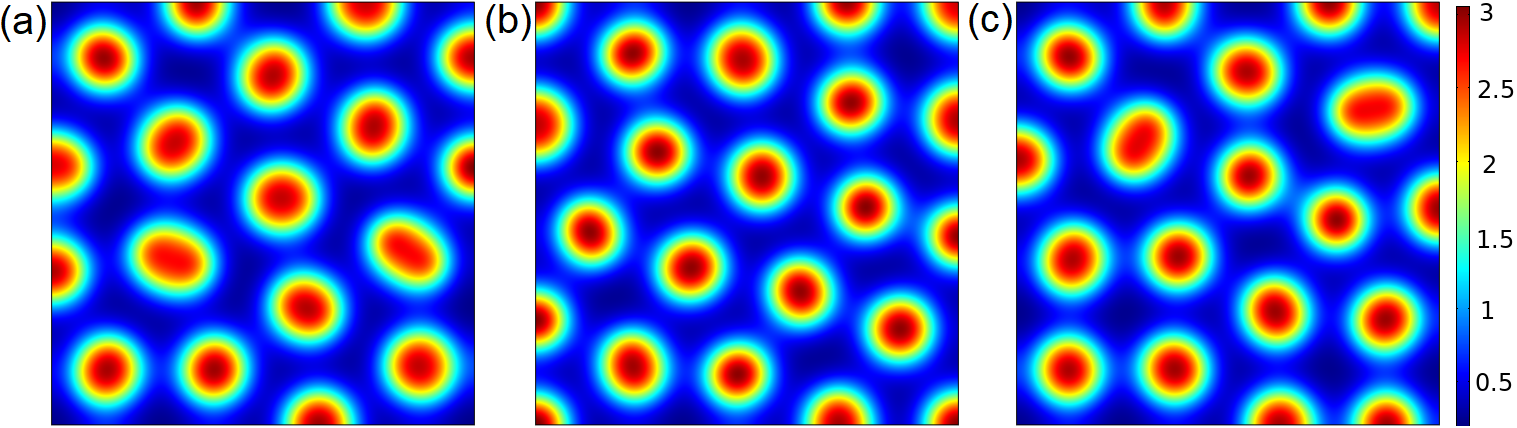}
			\vspace{-0.15in}
			\caption{Turing patterns corresponding to the function $u$ at $t=10000$ emergent from the Cattaneo system \eqref{Eq:Reaction-Telegraph_Equations} under the Schnakenberg kinetics $f(u,v) = \frac{1}{10}-u+u^2v$, $g(u,v) = 1-u^2v$, $d_{11}=0.0005$, $d_{22} = 0.01$ giving spot patterns. Simulations were considered on a unit square, as in Fig. \ref{Fig1}. We take (a) $\tau_1 =\tau_2 =0$, (b) $\tau_1=\tau_2=1$, (c) $\tau_1=0$, $\tau_2=5$.
				\label{Fig2}}
		\end{center}
	\end{figure}
	
	In Fig. \ref{Fig2} we give examples of patterns obtained on a planar domain. In these examples, and many others simulated (not shown), the impact of the relaxation time parameters, $\tau_1$ and $\tau_2$, was only transient, and did not seem to meaningfully change the final steady state patterns obtained, as was the case in the hyperbolic reaction-diffusion systems shown in Fig. \ref{Fig1}. This is again consistent with steady state solutions to these systems satisfying the same equation as the standard reaction-diffusion model, as well as with the influence of the inertial terms primarily affecting transient time evolution, but having no influence on the final patterned state which could be distinguished from different random initial conditions. We find similar results for other sets of kinetics.

	\subsection{Spatiotemporal patterns under the wave instability}\label{secBrusselator}
	We now illustrate the theory of Sec. \ref{THI} corresponding to the wave instability. Such an instability should, in principle, lead to spatiotemporal behavior due to spatial modes oscillating in time. While dispersion curves for such wave instabilities were computed in \cite{zemskov2016diffusive}, no simulations were shown to compare against the linear theory. Here we show an excellent match with the predictions from the linear theory for the initial evolution of a perturbation of the homogeneous equilibrium. We consider the hyperbolic reaction-diffusion system
	\begin{equation}
		\label{applicationgenwave}
		\begin{aligned}
			\tau_{1}\frac{\pa^2 u}{\pa t^2} + \frac{\pa u}{\pa t} & = d_{11}\nabla^2 u +5-10u +u^2 v,\\
			\tau_{2}\frac{\pa^2 v}{\pa t^2}  + \frac{\pa v}{\pa t} & = d_{22}\nabla^2 v +9u - u^2 v,
		\end{aligned}
	\end{equation}
	where we have taken Brusselator reaction kinetics.

	We remark that in all of the numerical simulations resulting in steady patterns, the systems were numerically stiff only for transient times. As solutions relaxed toward an inhomogeneous equilibrium, reasonably large timesteps could be safely taken. In contrast, the wave instabilities that we explored remained stiff for all times, as the systems never equilibrate to a steady solution but undergo continuous spatiotemporal oscillations. We therefore first considered simulations in one spatial dimension, which we implemented in MATLAB using the standard three-point centred-difference stencil for the Laplacian, and the stiff solver `ode15s' for the timestepping, with $N=500$ grid points. We took the absolute and relative tolerances to both be $10^{-11}$. As before, initial data were normally distributed random perturbations of the homogeneous equilibrium with standard deviation of $10^{-2}$. Simulations using COMSOL gave quantitative agreement for sufficiently strict tolerances.
	
\begin{figure}
		\begin{center}
			\includegraphics[width=1\textwidth]{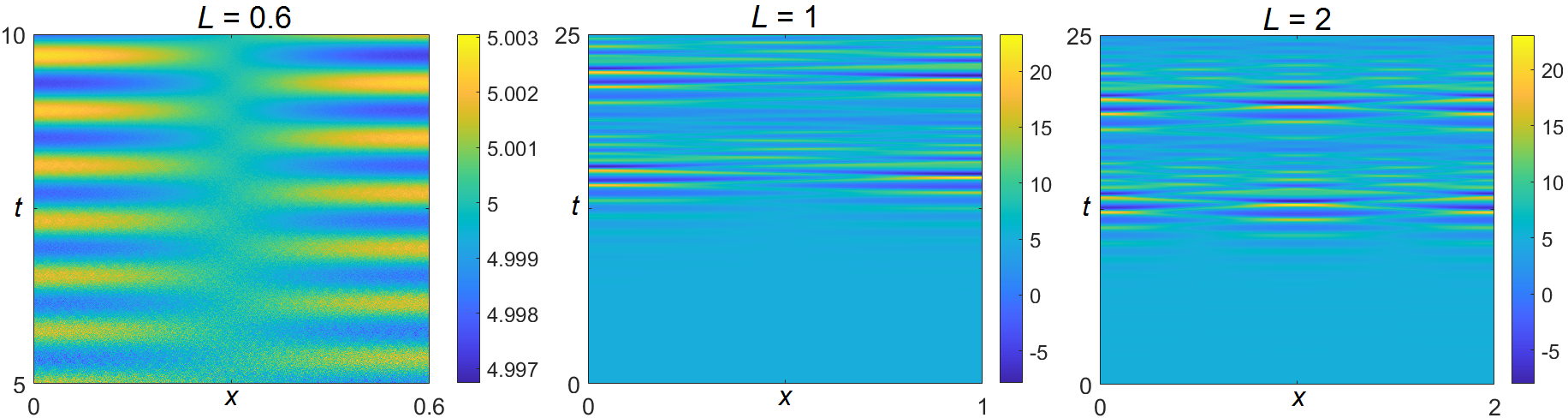}
			\vspace*{-0.25in}
			\caption{Simulations of the hyperbolic Brusselator system \eqref{applicationgenwave} on a 1-D domain of length $L$ as indicted. We took $d_{11}=3$, $d_{22}=1$, and $\tau_1=\tau_2=1$. \label{Fig3}}
		\end{center}
	\end{figure}

	For the parameters $\tau_1=\tau_2=1$, the analysis above gives us an unstable range of wavenumbers in the interval $\rho_k \in [1.917, 29.08]$. In the 1-D setting we have that $\rho_k = (k \pi/L)^2$, so we expect that random perturbations on domains of size $L \leq L^* = 0.58$ have no unstable wavemodes, but those on larger domains will have one or more. Simulations for values of $L<L^*$ confirm that random perturbations return to homogeneity, with small transient oscillations, whereas those on larger domains give rise to spatiotemporal oscillations. We give three examples of this in Fig. \ref{Fig3} on domains of size $L=0.6, 1,$ and $2$. In the first case, as the domain is just large enough to have an unstable $k=1$ mode, we anticipate that $\text{Re}(\lambda_1)$ is small and that the amplitude of deviations from the homogeneous steady state should be small. This is confirmed for $L=0.6$, where we see the $k=1$ mode oscillating around the steady state of $u^*=5$ with a small amplitude. For $L=1$, we see a much higher amplitude deviation from the steady state, with oscillations largely still following the $k=1$ mode in space though with slow changes in amplitude visible for $t>12.5$, indicative of multiple oscillation frequencies. Finally, when $L=2$, the modes $k=1,2,3$ are unstable, and observe that the $k=2$ mode seems to dominate for moderate timescales, with temporal oscillations around the spatial function $\cos(x\pi)$. For larger times we see a more complicated interplay of modes due to nonlinearities, but this example demonstrates that the instability theory of Sec. \ref{sectheory} correctly predicts the dominant mode initially driving oscillations away from the homogeneous state.
	
	Spatiotemporal structures obtained under the wave instability find perhaps their most natural state on manifolds without boundary, as there is no reflection due to no-flux boundary conditions. We now consider wave instability simulations on the surface of manifolds, but first make a comment on our confidence in the numerical methods used. We performed 1-D simulations, as in the above example, in COMSOL with $2,000$ finite elements, using both the BDF15 timestepping scheme and a generalized alpha scheme, both with a tolerance of $10^{-4}$. After checking convergence and setting minimum timesteps of $10^{-3}$ in all cases, we obtained quantitatively identical evolutions as in Fig. \ref{Fig3} to within $\sim 7$ decimal places of accuracy. In all subsequent simulations, we used the BDF timestepping scheme in COMSOL, but enforced a minimum order of 2 to ensure that high-frequency modes were not overdamped. We performed convergence checks in time by restricting the maximum timesteps to $10^{-2}$, and for shorter timescales $10^{-4}$, and observed comparable simulation profiles for all times, and quantitative agreement for timescales of order $20$ units. We anticipate that for longer timescales, mode mixing leads to sensitivity to initial conditions and hence the timestepping used will lead to quantitative disagreement.
	
	\begin{figure}[t!]
		\begin{center}
			\includegraphics[width=0.8\textwidth]{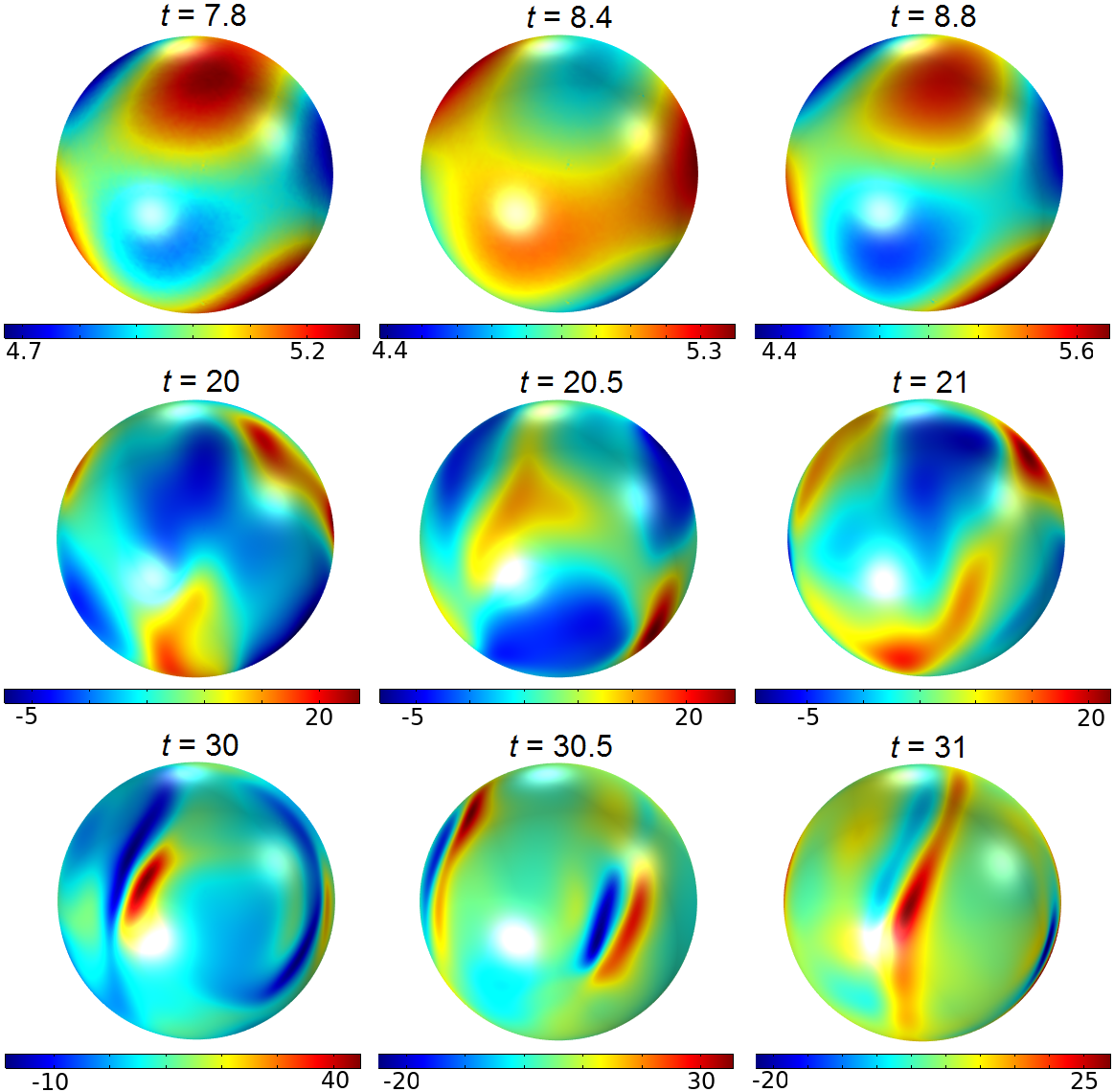}
			\caption{Simulations of the hyperbolic Brusselator system \eqref{applicationgenwave} on the surface of a sphere of radius $1$. We took $d_{11}=3$, $d_{22}=1$, and $\tau_1=\tau_2=1$. The equations were discretized using 48,788 triangular boundary elements. \label{Fig4}}
		\end{center}
	\end{figure}
	
	In Fig. \ref{Fig4} we give an example of the same system \eqref{applicationgenwave} simulated on the surface of a unit sphere. For short timescales ($t=7.8$ to $t=8.8$), we see oscillations largely around a single unstable mode, which we suspect matches the largest value of $\text{Re}(\lambda)$. For longer timescales $t=20$ to $t=21$, as in Fig. \ref{Fig3}(b,c), we observe more complicated spatiotemporal behavior which is no longer clearly an oscillation about a linear mode. Eventually this leads to a large increase in local amplitudes for $t>30$, and these structures rotate about the sphere in the form of highly-peaked waves. This rotating behavior persists for at least $20$ more units of time, with increasing amplitude.

	\begin{figure}[t!]
		\begin{center}
			\includegraphics[width=0.9\textwidth]{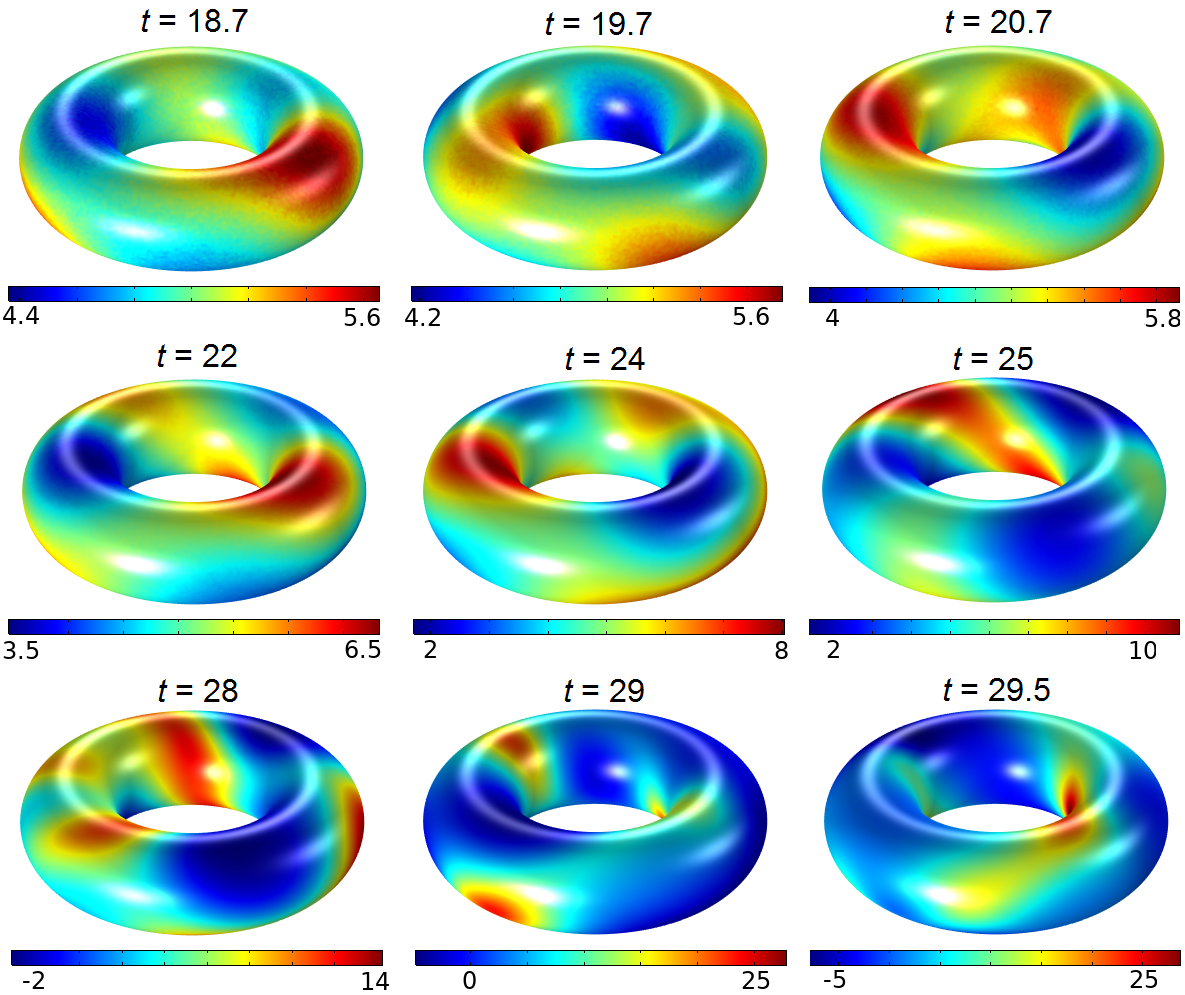}
			\caption{Simulations of the  hyperbolic Brusselator system \eqref{applicationgenwave} on the surface of a torus of major radius $1$ and minor radius $0.4$. Each panel corresponds to a fixed time, as indicated. We took $d_{11}=3$, $d_{22}=1$, and $\tau_1=\tau_2=1$. The equations were discretized using 58,882 triangular boundary elements. \label{Fig5}}
		\end{center}
	\end{figure}
	
	In Fig. \ref{Fig5}, we consider dynamics of the same system on the surface of a torus. Possibly due to the difference in surface area, it takes about twice as long for the initial noisy data to dissipate (see the small spatial noise for $t=18.7$ and $t=19.7$). We again see what appear to be close to pure mode oscillations up to $t\sim 20$, though there is some apparent drift around the torus. As time increases, this drift leads to a rotation of the spatial distribution around the torus as seen for $t=22$ to $t=25$, and finally to similarly highly localized peaks moving across the surface for $t>28$, as in the case of the sphere. We note that these snapshots only show the rotation of these peaks, but there is a simultaneous oscillation of the entire solution about the homogeneous steady state as in the 1-D case, leading overall to a quite complex evolution of the system.
	
	These simulations are qualitatively different from those seen in Turing-Hopf oscillations, such as those on the torus and the sphere in the FitzHugh-Nagumo system \cite{sanchez2019turing}. In that setting, whether or not spatiotemporal behavior was observed depended somewhat sensitively on both parameters and initial perturbations from the homogeneous steady state, as the temporal dynamics was a delicate balance between Turing and Hopf modes. In contrast, the wave instability explored in this section always leads to spatiotemporal dynamics, as long as the domain can accommodate an unstable wavemode and there are no other attractors. Additionally, the emergent dynamics here can clearly be seen to be dominated by a single spatial wavelength, at least for short time, which itself oscillates around the homogeneous steady state. In contrast, simulations of Turing-Hopf instabilities often involve many disparate lengthscales with wave-like movement of regions of high and low solution values, and appear to be spatiotemporally chaotic for even short timescales.

	\subsection{The role of cross-diffusion}
	
\begin{figure}[t!]
		\begin{center}
			\includegraphics[width=0.8\textwidth]{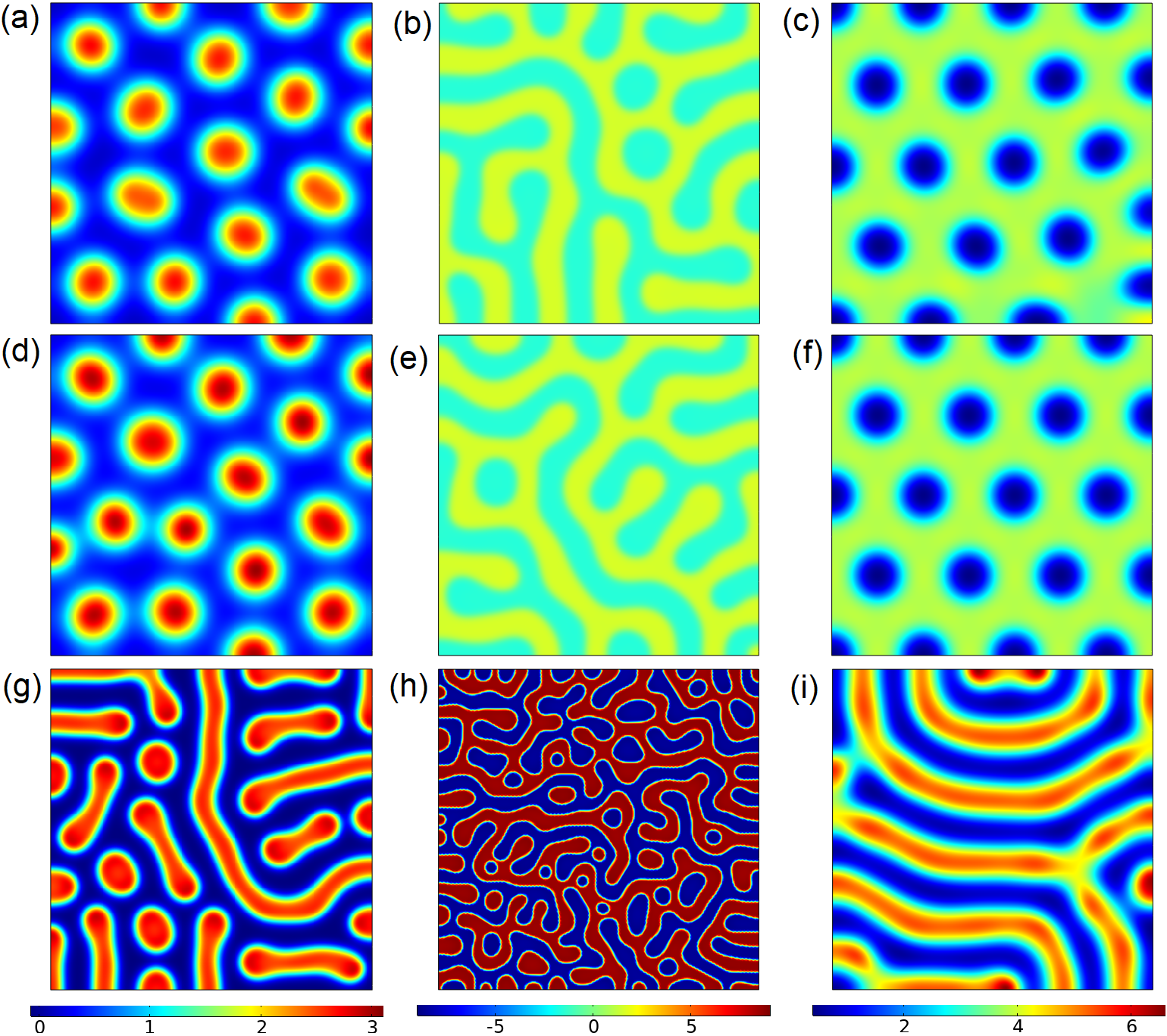}
			\caption{Turing patterns corresponding to the function $u$ at $t=10000$ emergent from reaction kinetics and diffusion parameters given by (a,d,g) the Schnakenberg model \eqref{schparams}, (b,e,h) the FitzHugh-Nagumo model \eqref{fnparams}, (c,f,i) the model giving sunken spots \eqref{sunkenparams}. In (a)-(c) we consider the standard Turing system \eqref{standard} with $\tau_1=\tau_2=F_{12}=F_{21}=d_{12}=d_{21}=0$, $F_{11}=F_{22}=1$ in each plot. We then vary parameters like (d) $\tau_1=\tau_2=0.1$, $F_{12}=F_{21}=0.5$, (e) $\tau_1=\tau_2=0.2$, $F_{12}=F_{21}=0.5$, (f) $\tau_1=\tau_2=0.2$, $F_{12}=F_{21}=1$, (g) $d_{12}=d_{21}=0.002$, (h) $d_{12}=0.1$, (i) $d_{12}=d_{21}=-0.001$. Each column uses the color scale shown at the bottom.  \label{Fig6}}
		\end{center}
	\end{figure}

	Extending \eqref{applicationgen} to include cross-diffusion parameters, we have
	\begin{subequations}\label{applicationCD}
		\begin{align}
			\tau_{1}\frac{\pa^2 u}{\pa t^2} + F_{11}\frac{\pa u}{\pa t} + F_{12}\frac{\pa v}{\pa t} & = d_{11}\nabla^2 u + d_{12}\nabla^2 v +f\left(u,v\right),\\
			\tau_{2}\frac{\pa^2 v}{\pa t^2} + F_{21}\frac{\pa u}{\pa t} + F_{22}\frac{\pa v}{\pa t} & = d_{21}\nabla^2 u +d_{22}\nabla^2 v +g\left(u,v\right).
		\end{align}
	\end{subequations}
	We now simulate the hyperbolic reaction-diffusion system \eqref{applicationCD} under the reaction kinetics discussed in Sec. \ref{secturingappl}, highlighting the role of cross-diffusion parameters $d_{12}$ and $d_{21}$ on the final pattern in Fig. \ref{Fig6}. In the standard reaction-diffusion case, these kinetics and parameters are all known to permit Turing patterns as shown in panels (a-c) of Fig. \ref{Fig6}. Using the same initial data (i.e.~the same random realization of the perturbations), we explored a variety of both the temporal parameters, $\tau_k$ and $F_{k\ell}$, as well as the cross-diffusion parameters, $d_{k\ell}$ within the feasibility regions predicted in Sec. \ref{sectheory}. Patterns found inside of these regions are shown in Fig. \ref{Fig6}(d-i). In panels (d-f) of Fig. \ref{Fig6} we see that the $\tau_k$ and $F_{k\ell}$ parameters had little effect on the qualitative kinds of patterns observed, and this was seen across a range of signs and magnitudes explored. While these patterns are quantitatively different from the reaction-diffusion patterns in panels (a-c) of Fig. \ref{Fig6}, we note that the differences are likely due to small changes in mode evolution (as can be seen in classical reaction-diffusion models due to different small random perturbations), and that the stability of the long-time patterned states is not influenced much, if at all, by these parameters. In contrast, the cross-diffusion examples shown in (g-i) of Fig. \ref{Fig6} demonstrate qualitatively different patterns, both in structure and amplitude, for each set of kinetics. This is expected as the cross-diffusion parameters enter into the conditions for linear instability, as well as the steady state equations, and hence undoubtedly enter into nonlinear selection effects for the resulting patterns. We do not show plots with other values of $\tau_k$ and $F_{k\ell}$, as they all appear qualitatively similar.
	
	\begin{figure}
		\begin{center}
			\includegraphics[width=0.9\textwidth]{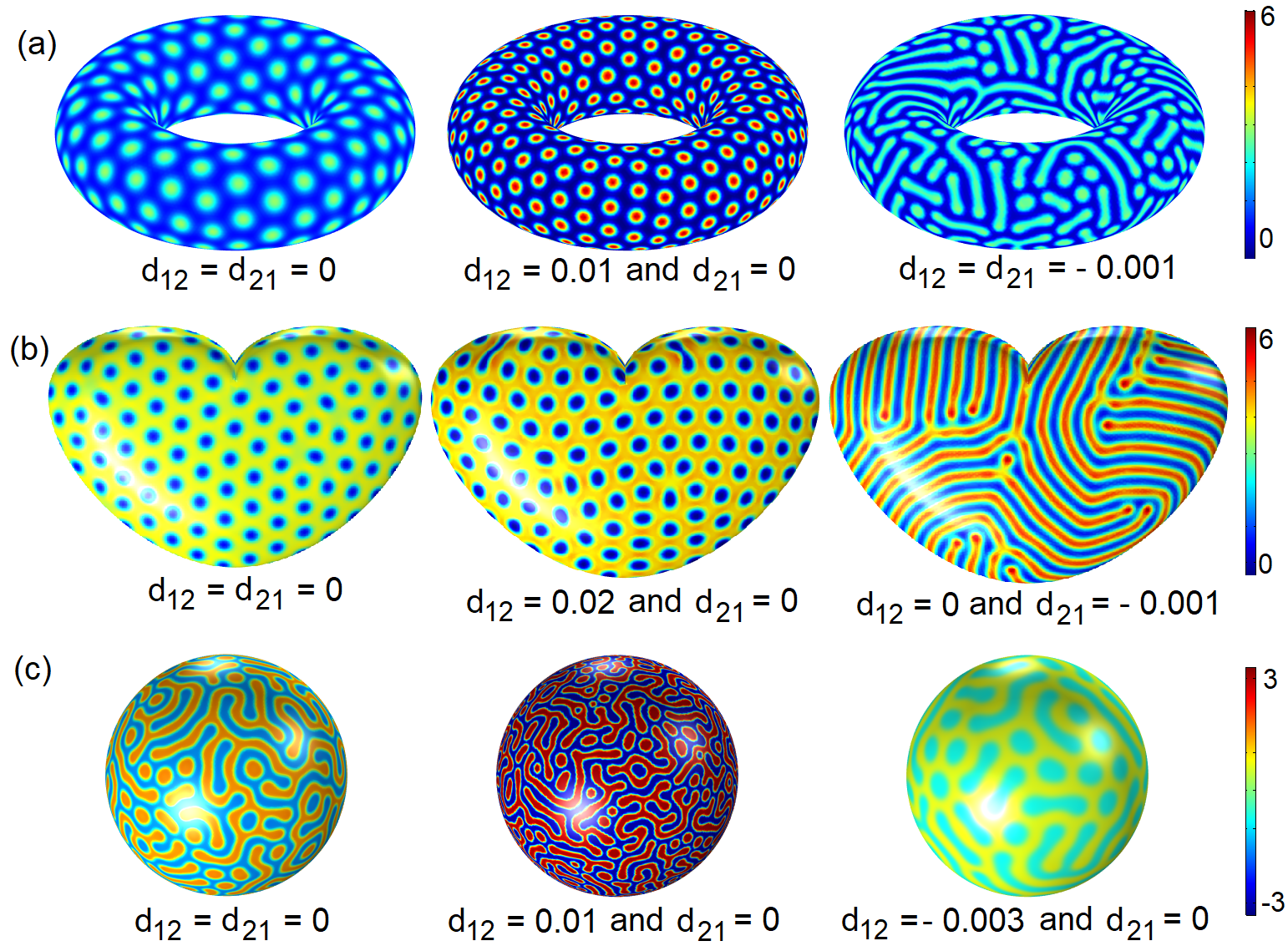}
			\caption{Surface Turing patterns corresponding to the function $u$ at $t=10000$ emergent from reaction kinetics and diffusion parameters given by (a) the Schnakenberg model \eqref{schparams} on a torus, (b) the model giving sunken spots \eqref{sunkenparams} on a heart-shaped surface, and (c) the FitzHugh Nagumo model \eqref{fnparams} on a sphere. In the first panel of each (left column) we consider the standard Turing system \eqref{standard}, and in the others we consider non-zero cross-diffusion terms. We take $\tau_1=\tau_2=F_{12}=F_{21}=0$, $F_{11}=F_{22}=1$ in each plot, while allowing the cross-diffusion parameters $d_{12}$ and $d_{21}$ to vary as indicated. Each row uses the color scale shown at the right. The torus has a minor radius of $0.4$, and a major radius of $1$, and is discretized by 41,266 triangular boundary elements. The heart-shaped domain is given parametrically by $x=\cos(\chi_1)\left (4\sqrt{1-\chi_2^2}\sin(|\chi_1|)^{|\chi_1|}\right )/2$, $y=\sin(\chi_1)\left (4\sqrt{1-\chi_2^2}\sin(|\chi_1|)^{|\chi_2|}\right )/2$, and $z=\chi_2/2$ for $\chi_1 \in [-\pi,\pi]$ and $\chi_2 \in [-1,1]$, and was discretized using 49,146 triangular boundary elements. The sphere was taken with radius $1$, and discretized with 76,580 triangular boundary elements. \label{Fig7}}
		\end{center}
	\end{figure}
	
	We also explored patterns on the surface of curved manifolds, to demonstrate the generality of the theory developed. In Fig. \ref{Fig7} we give three examples of patterns in the standard reaction-diffusion case corresponding to the first row of Fig. \ref{Fig6}, as well as how these patterns are modified by cross-diffusion. As in the planar example, cross-diffusion impacts both the wavelength and amplitude of patterns, as well as in some cases the qualitative structure, such as turning spots in Fig. \ref{Fig7}(a) or sunken spots in (b) into labyrinthine patterns under the right cross-diffusion parameters. In contrast, the parameters $\tau_k$ and $F_{k\ell}$ had no discernible impact on the resulting patterns within the feasibility regions, and so these simulations have been omitted. While the curvature plays some role in pattern selection, we have chosen sufficiently small wavelength patterns such that the local dynamics closely matches the same structures in the planar setting.
	
	The cross-diffusion terms appear to strongly modify the emergent patterns, while the temporal parameters associated to additional time derivatives do not. To understand the reason for this, assume there exists a final stationary pattern for the system \eqref{applicationCD}. Such a pattern corresponds to a time-independent solution $u=U(\mathbf{x})$, $v=V(\mathbf{x})$ satisfying the system
	\begin{subequations}\label{applicationCDstationary}
		\begin{align}
			0 & = d_{11}\nabla^2 U + d_{12}\nabla^2 V +f\left(U,V\right),\\
			0 & = d_{21}\nabla^2 U +d_{22}\nabla^2 V +g\left(U,V\right).
		\end{align}
	\end{subequations}
	As we see here, the self- and cross-diffusion terms enter into the stationary equations governing the pattern, while the time derivatives do not. The result is that, in the case of a stationary Turing pattern, the parameters associated with second-order time derivatives do not influence the final Turing pattern, rather they impact only the feasibility of said patterns. In contrast, the additional second-order space derivatives associated with cross-diffusion impact both the feasibility and final structure of the Turing pattern. They may also enlarge or contract the Turing space, depending on the parameter regime.
	
	\section{Discussion}\label{secdisc}
	We have studied Turing and wave instabilities from generic hyperbolic reaction-diffusion equations of the form \eqref{Eq:GeneralProblem}. We included hyperbolic structure in terms of second-order time derivatives, as well as self- and cross-diffusion, and generic first-order time derivatives accounting for self- or cross-friction. We obtained necessary and sufficient conditions for the linear Turing and wave instabilities in such systems, and then illustrated our theory with numerical simulations of the full nonlinear dynamics emergent from these systems.
	
	We find that the Turing instability is a generalization of the classical Turing instability in reaction-diffusion systems, and that the structure of the Turing conditions does not change compared to standard reaction-diffusion systems. The only difference is that the feasibility of the Turing instability is tied to additional conditions involving the additional temporal terms. In this way, some hyperbolic reaction-diffusion systems may permit the Turing instability whereas others will not, yet when the Turing instability is permitted, it takes the same form as found for classical reaction-diffusion systems with the same reaction kinetics and diffusion parameters. Furthermore, since Turing patterns correspond to a stable spatially heterogeneous steady state of \eqref{Eq:GeneralProblem}, the additional time derivatives do not modify the spatial structure of the pattern. This independence of stationary patterns on additional time derivatives contrasts the case of additional space derivatives, where past work has shown that resulting patterned structures are sensitive to additional second-order cross-diffusion terms \cite{vanag2009cross, gambino2013pattern} and first-order advection terms \cite{van2019diffusive}.
	
	The wave instability is not possible in classical systems of two reaction-diffusion equations, yet the addition of hyperbolic terms (second derivatives) allows for this instability to occur in the generic system we study. As the wave instability results in spatiotemporal rather than steady state spatially heterogeneous patterns, the time evolution of the system \eqref{Eq:GeneralProblem} needs to be retained in order to understand resulting patterns. This also means that the additional temporal parameters in \eqref{Eq:GeneralProblem} play a strong role in influencing the structure of any emergent spatiotemporal pattern, in addition to providing necessary and sufficient conditions for the existence of such a pattern. The Turing-wave instability also gives routes to symmetry breaking and pattern formation inadmissible for classical Turing instabilities, such as when the activator diffuses more quickly than the inhibitor. These mechanisms could help alleviate some of the restrictions typically found in Turing-type symmetry breaking, though we leave further investigation of this to future work. 
	
	Dynamics from the wave instability are qualitatively different from those seen in Turing-Hopf oscillations \cite{sanchez2019turing}. In that setting, the emergence of particular spatiotemporal behavior depended sensitively on both parameters and initial perturbations from the homogeneous steady state, as the temporal dynamics was a delicate balance between Turing and Hopf modes. In contrast, the wave instability explored in this paper always leads to spatiotemporal dynamics, as long as the domain can accommodate an unstable wavemode and there are no other attractors. Additionally, the emergent dynamics here can clearly be seen to be dominated by a single spatial wavelength, at least for short time, which itself oscillates around the homogeneous steady state. In contrast, simulations of Turing-Hopf instabilities often involve many disparate lengthscales with wave-like movement of regions of high and low solution values, and appear to be spatiotemporally chaotic for even short timescales. This suggests that the wave instability is a useful and robust mechanism for generating spatiotemporal patterns. Regarding other future work, Turing-Hopf bifurcations occur when a Turing instability takes place in the presence of a Hopf bifurcation of the spatially uniform steady state \cite{rovinsky1992interaction, meixner1997generic}. It may be useful to consider when a Hopf bifurcation of the steady state is coincident with a wave instability in the spatially extended system, and including hyperbolic terms would allow for this possibility within a system of two equations. 
	
	While continuum domains were considered in the present paper, it is also possible to consider diffusive instabilities in reaction-diffusion systems on discrete, lattice, or network domains \cite{nakao2010turing, wolfrum2012turing, mimar2019turing, muolo2019patterns, van2021theory}. This was done for the direct analogue of \eqref{oldpaper} for distinct inertial times by \cite{carletti2021turing}, although the wave instability was not classified in their work (all instabilities were treated as a type of Turing instability). Due to the finite dimensional nature of networks, some of the feasibility conditions in the analysis of Sec. \ref{sectheory} (in particular, those for $\rho_\ell \rightarrow \infty$) may be relaxed, and hence there may be more degrees of freedom for pattern selection. For this reason, an analysis of the wave instability on network domains would also be of interest. 
	
	\section{Conclusions}\label{secconclusions}
	Having studied generic routes to diffusive instabilities in systems of two hyperbolic reaction-diffusion equations, a few important conclusions emerge from our analysis:
	\begin{itemize}
		\item Second-order time derivatives only influence the feasibility regions and not the conditions for the Turing instability. In particular, since there are extra feasibility conditions yet the standard Turing conditions \eqref{TuringSC1}-\eqref{TuringSpec1} remain the same, second-order time derivatives will at most shrink the region of parameter space which supports Turing patterns. The final stationary Turing pattern is unchanged with the inclusion of second-order time derivatives.
		\item Second-order time derivatives allow for the possibility of the wave instability in a system of two hyperbolic reaction-diffusion equations. The wave instability is not possible for a system of two standard reaction-diffusion equations. 
		\item Additional second-order spatial derivatives, in the form of cross-diffusion terms, allow for a change in both the feasibility conditions and the instability conditions for both the Turing and wave instabilities. Furthermore, such terms can alter the form of the stationary or spatiotemporal pattern which develops. 
	\end{itemize}
	Therefore, if one is interested in permitting spatiotemporal pattern formation in a system of two reaction-diffusion equations, one might incorporate second-order time derivatives. On the other hand, if one is interested in changing the actual shape or structure of any emergent patterns, one should incorporate additional second-order space derivatives through cross-diffusion terms, but not additional time derivatives.
	
	We also considered a variety of degenerate cases modelled by simpler systems, with the following findings apparent from our analysis:
	\begin{itemize}
		\item Two distinct species are required for the diffusive instability resulting in pattern formation. A scalar hyperbolic reaction-diffusion equate does not admit diffusive instability, despite the additional second-order derivative. 
		\item Mixed hyperbolic-parabolic reaction-diffusion systems permit both the Turing and wave instabilities. These can therefore be seen as the minimal models (in terms of time derivatives) for two-species which admit both instabilities.
		\item Parabolic-parabolic or hyperbolic-elliptic systems may admit the Turing instability but not the wave instability.
	\end{itemize}
	These finding should be of use to those interested in designing reaction-diffusion systems to model pattern formation in real-world scenarios \cite{lengyel1992chemical, vittadello2021turing}. In particular, if one is interested in moving patterns akin to those found in the presence of the wave instability, then one needs both two species and at least three time derivatives. If one is interesting in stationary Turing patterns, a wider variety of two-species systems are compliant, with only two time derivatives needed. 
	
	\newpage
	\bibliographystyle{unsrt} 
	\bibliography{bibliography}
	
\end{document}